\documentclass[aps,pra,twocolumn,superscriptaddress,showpacs,aps]{revtex4-1}
\usepackage{amsmath,epsfig,amsfonts,graphicx}
\usepackage{subfigure}
\usepackage{graphicx}
\usepackage[american]{babel}
\usepackage{amssymb}
\usepackage{amsfonts}
\usepackage{color}
\usepackage[normalem]{ulem}
\usepackage{hyperref}
\begin{document}
\newcommand{\be}{\begin{equation}}
\newcommand{\ee}{\end{equation}}
\newtheorem{corollary}{Corollary}[section]
\newtheorem{remark}{Remark}[section]
\newtheorem{definition}{Definition}[section]
\newtheorem{theorem}{Theorem}[section]
\newtheorem{proposition}{Proposition}[section]
\newtheorem{lemma}{Lemma}[section]
\newtheorem{help1}{Example}[section]
\renewcommand{\theequation}{\arabic{section}.\arabic{equation}}
\title{Self-trapping transition for a nonlinear impurity within a linear
chain}
\author{Haitian Yue}
\affiliation{Department of Mathematics and Statistics, University of Massachusetts, Amherst MA 01003-4515, USA}
\author{Mario I. Molina}
\affiliation{Departamento de F\'{\i}sica, MSI-Nucleus on Advanced Optics,  and Center for Optics and Photonics (CEFOP), Facultad de Ciencias, Universidad de Chile, Santiago, Chile.}
\author{Panayotis G. Kevrekidis}
\affiliation{Department of Mathematics and Statistics, University of Massachusetts, Amherst MA 01003-4515, USA}
\author{Nikos I. Karachalios}
\affiliation{Department of Mathematics, University of the Aegean, Karlovassi, 83200 Samos, Greece}

\begin{abstract}
In the present work we revisit the issue of the self-trapping dynamical
transition at a nonlinear impurity embedded in an otherwise linear lattice.
For our Schr{\"o}dinger chain example, we present  rigorous arguments
that establish necessary conditions and corresponding
parametric bounds for the transition between linear decay
and nonlinear persistence of a defect mode. The proofs combine a contraction
mapping approach applied in the fully dynamical problem in the case of a $3$D-lattice, together with variational arguments for the derivation of parametric bounds for the creation of stationary states associated with the expected fate of the self-trapping dynamical transition. The results are relevant for both power law nonlinearities and
saturable ones. The analytical results are corroborated by numerical
computations. 
\end{abstract}


\maketitle

\section{Introduction}

The theme of discrete linear chains with embedded nonlinear
impurity nodes is one of considerable interest within condensed
matter physics. It emerges, for instance, within tight-binding descriptions of
electron transport, where the nonlinear terms describe local interactions
with vibrations at the impurity node~\cite{tsir1}, \cite{mol1}. It also
arises in the study of tunneling through a magnetic impurity 
connected to two perfect leads in the
presence of a magnetic field~\cite{moli}.
It is also fairly widespread in the realm of nonlinear optics, where
waveguides with practically linear and ones with essentially
nonlinear characteristics can be constructed. This was proposed e.g. 
in~\cite{efrem}; however, notice that in that context
linear and nonlinear waveguides were proposed to be interlaced in
binary arrays. Here, instead, we have in mind a single nonlinear waveguide
embedded in an otherwise linear array. Given this diverse
array of physical setups, this subject was numerically examined
in a wide array of studies~\cite{delyon,souk,molina1,molina2,molina3}, not
only for the case of one but also for that of more embedded impurities.
This topic has also recently seen a surge of renewed interest,
due in part to the examination of eigenvalue 
problems and symmetry-breaking features
in the presence of multiple linear or nonlinear impurity 
sites, and also due to the
examination of gain/loss variants thereof~\cite{malas1,malas2,jennie,jennie2,malas3}.

While numerically the relevant dynamics is rather straightforward and
directly tractable, on the analytical side, unfortunately, developments
have been considerably less advanced. While it is possible to characterize
the stationary states of the problem via the Green's function 
techniques~\cite{molina1}, and even (for specially constructed potentials)
to capture symmetry-breaking effects via the demonstration of emergence
of asymmetric states~\cite{malas1}, little has been rigorously
established about the
dynamic problem. Assuming that initially, the excitation is placed at  site $n=0$ of the chain, that is $C_n(0)=\delta_{n,0}$, the work of~\cite{molina2} was intriguing 
in that it established a particular diagnostic
\begin{eqnarray}
\langle P \rangle \equiv \lim_{t \rightarrow \infty}\frac{1}{t} \int_0^t |C_0(s)|^2 ds,
\label{eq0}
\end{eqnarray}
with a clearly distinct behavior for different parameters (i.e., nonlinearity
strengths $\chi$) of the system. The numerical (or physical) experiment
at hand is as follows. Suppose we initialize the nonlinear site at
unit intensity (the relevant amplitude can always be rescaled so that
there is one parameter, either the strength of the nonlinearity or
interchangeably the magnitude of the compactly supported --on a single
site-- initial data). We then monitor $\langle P \rangle$ [in fact, in
our case, we will not compute the relevant integral numerically starting
from
$t=0$ but rather from $t=25$ to exclude short term transient dynamics].
We then observe that for $\chi < \chi_c$ (which for the cubic nonlinear
case is $\chi_c \approx 3.2$), our diagnostic quantity tends to 0.
On the contrary, for $\chi > \chi_c$, the quantity remains finite
and displays an increasing trend as $\chi$ increases~\cite{molina2}.

In the present work, we initiate our investigations for establishing such a
behavior from a rigorous perspective, and examine (both analytically
and numerically) how the behavior is modified for different types of
nonlinearities. In particular, we will focus on power law nonlinearities,
as well as on saturable ones. 

More specifically, we will 
consider the discrete nonlinear Schr\"{o}dinger equation (DNLS), with the single nonlinear impurity in a $N$-dimensional lattice, $N\geq 1$,
\begin{eqnarray}
\mathrm{i}\frac{d C_{n}}{dt}=\sum_{m}V_{n,m}C_{m}-\chi\delta_{n,0}F(C_n),\;\;n\in\mathbb{Z}^N.\;\;\;
\label{eq:1}
\end{eqnarray}
where $n=(n_1,n_2,\ldots,n_N)$, and $V_{n,m}$ is the coupling between sites $n$ and $m$.  

Motivated by applications in the context of nonlinear 
optics~\cite{dnls_book}, we shall concentrate 
on two examples of nonlinearities $F:\mathbb{C}\rightarrow\mathbb{C}$, 
namely the power-law  nonlinearity
\begin{eqnarray}
\label{nonl1}
F(z)=|z|^{2\sigma}z,
\end{eqnarray}
and the saturable nonlinearity
\begin{eqnarray}
\label{nonl2}
F(z)=\frac{|z|^2 z}{1+|z|^2},
\end{eqnarray}
with the single site, unit intensity initial condition  
\begin{eqnarray}
\label{eqin1}
C_n(0)=\delta_{n,0},
\end{eqnarray} 
indicated above. Note that here $C_0(t)$ denotes $C_{(0,...,0)}(t)$ in the $N$-dimensional lattice.

We follow two different approaches: The first approach is based on the study of a nonlinear integral equation for $C_0(t)$, derived in section \ref{SECII}. However, in this dynamical approach, we shall use a different diagnostic for self-trapping than (\ref{eq0}), namely 
\begin{eqnarray}
\label{eq0a}
\lim_{t\rightarrow\infty}|C_0(t)|^2.
\end{eqnarray}

Then, in section \ref{SECIII}, and in the case of the $3$-dimensional lattice, we apply on the integral equation a contraction mapping approach, and prove the following: there exists a critical value $\chi_{\mathrm{crit}}$, such that if
$\chi <\chi_{\mathrm{crit}} $, then
$\lim_{t\rightarrow\infty}|C_0(t)|^2=0$.  
Although our rigorous 
result is restricted to the $3$D-case, it is also suggestive
of the numerically identified dynamical transition. Furthermore, it appears that the corresponding value for the saturable nonlinearity is greater than the one for the power nonlinearity.  Next, by using an approximation near the linear regime
of the system, we will rationalize an increasing behavior of the possible critical point $\chi_{\mathrm{crit}}$ for the power nonlinearity,
as well as, unify our critical point 
estimates for the cubic and saturable case. We remark that the restriction in the dimension $N=3$ is imposed by integral divergence properties of the kernel of the integral map.

In section \ref{SECIV} we discuss our second, variational approach. 
The latter is motivated by numerical results, illustrating that when $\chi$ is greater than the actual numerical value $\chi_{c}$, $|C_0(t)|$ is approaching a 
non-vanishing constant value. Such a convergence suggests that in the 
strongly nonlinear regime $\chi>\chi_{c}$, and after a long time, that the
dynamics of $C_0(t)$ is approaching a stationary (i.e., standing
wave) state. This, in turn, motivates us to discuss the possible existence of critical values on $\chi$ for the formation of stationary states. On the one hand, and under the light of the above assumption on the long time asymptotics, we work on the integral equation and derive critical values for the $1$D and $2$D cases, for both types of nonlinearities. On the other hand, the  suggested dependence of the derived critical values on the dimension $N=1,2$, motivates us to
apply constrained minimization arguments on the stationary DNLS equation in the lattice $\mathbb{Z}^N$, $N\geq 1$.   Interestingly, the obtained critical values on $\chi$ 
(for the existence of such standing waves) both
unify and extend the observations from the manipulation of the integral equation in the $1$D and $2$D cases, and demonstrate the following qualitative features:  (a) In the case of the power 
nonlinearity, the critical value increases, as the nonlinearity exponent 
$\sigma$ increases. (b) Both critical values for power and saturable nonlinearities, increase as the dimension of the lattice $N$ increases. (c) The critical value for the saturable nonlinearity is larger than the critical value for the cubic power nonlinearity. 

In Section \ref{SECV}, we present the numerical computations of the physical quantity $\langle P \rangle$, as well as for (\ref{eq0a}), and examine the self-trapping dynamical transition. The computations have been performed for  lattice dimensions $1\leq N\leq 3$. 
In all cases, the numerical results confirm the existence of a
dynamical transition.
Furthermore, the numerical critical values  $\chi_c$ for self-trapping share 
the above analytical trends, such as the dependence 
on the nonlinearity exponent, the type of nonlinearity, as well as, the 
dimension of the lattice with the critical points for the existence   
of stationary states that we present in section~\ref{SECIV}.

Finally, in section VI, we will summarize our findings and
present our conclusions, as well as suggest some potential
directions for future work.

The complementary section \ref{AppPa} includes for completeness, some basic properties of the eigenvalue problem for the discrete Laplacian on $\mathbb{Z}^N$, supplemented with Dirichlet boundary conditions, while section~\ref{AppPb} 
contains some special function properties of use to our stationary
state calculations.

We should stress here that there are a few features that distinguish
our results from earlier works discussed above that identified the
full stationary family of solutions to the 
problem of a single defect (and even to that of multiple defects)
within a linear lattice and that also numerically demonstrated the
dynamical self-trapping transition, as discussed above. 
 Here, not only is the self-trapping transition explored
numerically for a variety of models (power or saturable nonlinearities)
and a variety of dimensions (1D, 2D and even 3D) but importantly 
the transition is rigorously proven to exist in sufficiently high
dimension (N=3) and additional arguments are given approximating its
critical point in the different cases examined.

\section{Derivation of an integral equation for $C_0(t)$}
\label{SECII}
\setcounter{equation}{0}
In this section we will derive a dynamical, nonlinear integral equation for the evolution of $C_0(t)$. 
Let us consider first the case $N=3$.  We define the Fourier transforms
\begin{eqnarray}
C_{ k}&=& \sum_{ n} C_{ n} \exp(\mathrm{i} { k}\cdot{ n})
\label{eq:2a}\\
C_{ n}&=& \int {d^3 k\over{(2 \pi)^3}} C_{ k} \exp(-\mathrm{i} { k}\cdot{ n})\label{eq:2}
\end{eqnarray}
where ${ k}=(k_{1},k_{2},k_{3})$.
Taking the Fourier transform  (\ref{eq:2a}) of Eq.~(\ref{eq:1}), and assuming $V_{{ n},{ m}}=V_{ {n-m}}$,  we get the equation
\be
\mathrm{i} {d C_{ k}\over{d t}}= \lambda_{ k} C_{ k}- \chi F (C_{ 0}),\label{eq:3}
\ee
where
\be
\lambda_{ k}=\sum_{ n} V_{{ n}-{ m}}\exp(\mathrm{i} { k}\cdot({ n}-{ m})),
\ee
is the dispersion relation. Now we take the Laplace transform of Eq.(\ref{eq:3})
\be
\mathrm{i} (\omega \widetilde{C_{ k}}-1)=\lambda_{ k} \widetilde{C_{ k}} - \chi \widetilde{F}(\omega),
\ee
that is,
\be
\widetilde{C_{ k}}={1\over{\omega + \mathrm{i} \lambda_{ k}}} + \mathrm{i} {\chi \widetilde{F}(\omega)\over{\omega + \mathrm{i} \lambda_{ k}}}.\label{eq:6}
\ee
Then, we take the inverse Laplace transform of Eq.~(\ref{eq:6}). The first term is the Laplace  transform of $\exp(-i \lambda_{ k}t)$. The second term is a product of two transforms, and therefore, its inverse is a convolution of $\exp(-i \lambda_{ k}t)$ and $F(C_{0}(t))$. Hence,
\begin{eqnarray}
C_{ k} &=& \exp(-\mathrm{i} \lambda_{ k}t)\nonumber\\ 
&+& \mathrm{i} \chi \int_{0}^{t} \exp(-\mathrm{i} \lambda_{ k}(t-s)) F(C_{0}(s)) ds
\label{eq:7}
\end{eqnarray}
Up to this point, the treatment is completely general. To obtain a specific result, we must use the exact form of the dispersion relation $\lambda_{ k}$. In our case,
 $\lambda_{ k}=2 V (\cos(k_{1})+\cos(k_{2})+\cos(k_{3})$.
The last step is to take the inverse Fourier transform 
of Eq.~(\ref{eq:7}). Using Eq.~(\ref{eq:2}), the first term will be
\be
\int {d^3 k \over{(2 \pi)^3}} \exp[-\mathrm{i} ( \lambda_{ k}t + { k}\cdot{ n})],
\ee
that is,
\begin{eqnarray}
&&\int {dk_{1}\over{2 \pi}}e^{[-\mathrm{i}2V t \cos(k_{1})-\mathrm{i} k_{1}n_{1}]}\int {dk_{2}\over{2 \pi}}e^{[-\mathrm{i}2V t \cos(k_{2})-\mathrm{i}k_{2}n_{2}]}\nonumber\\
&&\times\int {dk_{3}\over{2 \pi}}e^{[-\mathrm{i}2V t \cos(k_{3})-\mathrm{i} k_{3}n_{3}]}.
\end{eqnarray}
Using the integral representation of the Bessel functions
\be
\int_{-\pi}^{\pi} {dk\over{2 \pi}} e^{-\mathrm{i} (k n + 2 V t \cos(k))} = (-\mathrm{i})^n J_{n}(2 V t),
\ee
the first term in Eq.~(\ref{eq:7}) leads to $(-\mathrm{i})^{n_{1}+n_{2}+n_{3}} J_{n_{1}}(2 V t) J_{n_{2}}(2 V t) J_{n_{3}}(2 V t)$.

For the second term in Eq.(\ref{eq:7}), we exchange the integrals over ${ k}$ and $s$ and arriving to
\be
\mathrm{i} \chi \int_{0}^{t} F(C_{0}(s)) \times 
\prod_{j=1}^{3}(-\mathrm{i})^{n_{j}} J_{n_{j}}(2 V (t-s))ds.
\ee
Finally, putting everything together, we derive
\begin{eqnarray}
C_{ n}(t) &=& \prod_{j=1}^{3}(-\mathrm{i})^{n_{j}} J_{n_{j}}(2 V t)\nonumber\\ 
&+& \mathrm{i} \chi \int_{0}^{t} F(C_{0}(t))\prod_{j=1}^{3}(-i)^{n_{j}} J_{n_{j}}(2 V (t-s)).\;\;\;\;\;\;\;\;\;\label{eq:12}
\end{eqnarray}

In particular, for $n=(0,0,0)$ we have
\begin{eqnarray}
C_{0}(t) &=& J_{0}(2 V t)^3\nonumber\\ 
&+& \mathrm{i} \chi \int_{0}^{t} J_{0}[2 V (t-s)]^3 F(C_{0}(s)) ds,
\label{eq:13}
\end{eqnarray}
or, equivalently,
\begin{eqnarray*}
C_{0}(t) &=& J_{0}(2 V t)^3\nonumber\\ 
&+& \mathrm{i}\chi \int_{0}^{t} J_{0}(2 V s)^3 \ \ F[C_{0}(t-s)] ds.
\end{eqnarray*}

For the $N$-dimensional cubic lattice, the extension of the  formulas (\ref{eq:12}) and (\ref{eq:13}) is rather clear:
\begin{eqnarray*}
C_{ n}(t) &=& \prod_{j=1}^{N}(-\mathrm{i})^{n_{j}} J_{n_{j}}(2 V t)\nonumber\\ 
&+& \mathrm{i}\chi \int_{0}^{t} F(C_{0}(t))\prod_{j=1}^{N}(-\mathrm{i})^{n_{j}} J_{n_{j}}(2 V (t-s)).
\end{eqnarray*}
and 
\begin{eqnarray}
C_{0}(t) &=& J_{0}(2 V t)^N\nonumber\\ 
&+& \mathrm{i} \chi \int_{0}^{t} J_{0}[2 V(t-s)]^N \ \ F(C_{0}(s)) ds.
\label{eq:15}
\end{eqnarray}

We are interested in determining the presence or absence of a {\em self-trapping transition} in the sense of the diagnostic (\ref{eq0a}), i.e., to determine if there is a critical
value $\chi_c$ for which
\begin{equation}
\label{eq4}
\lim_{t\rightarrow\infty}|C_0(t)|^2=\left\{
\begin{array}{lll}
&0,&\;\;\mbox{if}\;\;\chi<\chi_{c},\\
&\neq0,&\;\;\mbox{if}\;\;\chi>\chi_{c}.
\end{array}
\right.
\end{equation}

As a side note, we should point out that the effective integrability
of the linear lattice problem with the defect (as manifested e.g.
in Eq.~(\ref{eq:15})) is a starting point for our analysis of
the contractive dynamics. We now turn to the rigorous study
of the latter for $N=3$.

\section{Contractive Dynamics for the integral equation in the case $N=3$.}
\setcounter{equation}{0}
\label{SECIII}
It is well known that the solutions of the DNLS equation  (\ref{eq:1}) exist globally in time (e.g. are of class $C([0,\infty),\ell^2)$, where $\ell^2$ denotes the space of square summable sequences. One of the two principal
conserved quantities is the {\em power} or (squared $l^2$) 
{\em norm} $\mathcal{P}=\sum_{n\in\mathbb{Z}^N}|C_n(t)|^2$, and 
for all $t\geq 0$,
\begin{eqnarray}
\label{eq5a}
\sum_{n\in\mathbb{Z}^N}|C_n(t)|^2=\sum_{n\in\mathbb{Z}^N}|C_n(0)|^2.
\end{eqnarray}
Then for the initial condition considered herein, 
the corresponding unique solution of the DNLS equation (\ref{eq:1}) satisfies 
\begin{eqnarray}
\label{eq5}
\sum_{n\in\mathbb{Z}^N}|C_n(t)|^2=1,\;\;\mbox{for all}\;\;t\geq 0.
\end{eqnarray}
The norm conservation (\ref{eq5}) implies that $C_0(t)$ satisfies the uniform in time estimate
\begin{eqnarray}
\label{eq6}
|C_0(t)|\leq 1,\;\;\mbox{for all}\;\;t\geq 0.
\end{eqnarray}

It is evident from (\ref{eq6}) that 
\begin{eqnarray}
\label{ena}
\langle P \rangle\leq 1.
\end{eqnarray}
Due to (\ref{eq6}), we have $\frac{1}{t}\int_0^t|C_0(s)|^2ds\leq \frac{1}{t}\int_0^tds=1$, for all $t\geq 0$. Then, by passing to the limit as $t\rightarrow\infty$, we get (\ref{ena}).

The observed phenomenology of the self-trapping transition as discussed in \cite{molina2} is that in the regime $\chi\geq\chi_c$, there is a sudden increase of the time-averaged probability $\langle P \rangle$, while in the regime $\chi<\chi_c$,  the average $\langle P \rangle=0$.  This phenomenology suggests to investigate the contracting dynamics of the integral equation
(\ref{eq:15}) in an appropriate regime for $\chi$. Having assigned through the initial condition the initial value $C_0(0)=1$ at $t=0$, and since $C_0(t)$ is uniformly bounded from (\ref{eq6}), it is convenient  mathematically 
to consider the self-trapping transition problem in the Banach space of the {\em essentially bounded functions} $L^{\infty}([0, \infty))$ endowed with the norm $||u||_{\infty}=\mathrm{ess\,sup}_{[0,\infty))}|u(t)|$, \cite{Adams}. 

The results that will be presented in this section are restricted to the $3\mathrm{D}$-lattice. In the case $N=3$, 
the integral equation (\ref{eq:15}) will define a nonlinear map $\mathcal{T}:L^{\infty}([0,T])\rightarrow L^{\infty}([0,T])$, for arbitrary large $0<T<\infty$,  by the equation 
\begin{eqnarray}
\label{eq8}
\mathcal{T}[C_0(t)]&=& J_0(2Vt)^3\nonumber\\
&+&\mathrm{i}\chi\int_{0}^{t}J_0[2V(t-s)]^3F(C_0(s))ds.
\end{eqnarray}
For instance, due to (\ref{eq6}) we shall consider the nonlinear map $\mathcal{T}$ on the closed unit ball of $L^{\infty}([0,T))$
$$B=\left\{u\in L^{\infty}([0,T])\;\;:\;\;||u||_{\infty}\leq 1+\varrho\right\},$$
where $\varrho$ is a constant which will be fixed later on. Then, we will seek for the existence of a regime of the parameter $\chi$ so that $\mathcal{T}:B\rightarrow B$ will be a contraction, and examine the behavior of the unique fixed point for large times. 
\begin{theorem}
\label{th1}
Consider the DNLS system (\ref{eq:1}) with the nonlinearities (\ref{nonl1}) or (\ref{nonl2}) in the case $N=3$. (a) There exists a critical value $\chi_{\mathrm{crit}}$ such that if $\chi<\chi_{\mathrm{crit}}$, then $\lim_{t\rightarrow\infty}|C_0(t)|=0$. (b) The critical value for the power nonlinearity is smaller than the critical value for the saturable one.
\end{theorem}
\textbf{Proof:} 
We distinguish between the cases of the power (\ref{nonl1}) and saturable nonlinearity (\ref{nonl2}) respectively. 
\newline
\emph{A. Power nonlinearity.} First, we will examine the conditions under which $\mathcal{T} : B \to B,$ i.e., $\mathcal{T}$ maps B to itself. For any $C_0\in B$, we observe that
\begin{eqnarray}
 |\mathcal{T}[C_0(t)]|
&\leq& |{J_0(2Vt)}|^3\nonumber\\ 
&&+ \chi \int_0^t|{J_0[2V(t-s)]}|^3 |F(C_0(s))|ds\nonumber\\
&\leq& 1 + \chi (1+\varrho)^{(1+2\sigma)} \int_0^t|{J_0[2V(t-s)]}|^3 ds\nonumber\\
&=& 1 +  \chi(1+\varrho)^{(1+2\sigma)} A(t),
\label{eq:H1}
\end{eqnarray}
where the function $A(t)$ is defined as
\begin{eqnarray*}
A(t)=\int_{0}^{t}\left|J_0[2V(t-s)]\right|^3 dy.
\end{eqnarray*}
Note that the function $A\in L^{\infty}([0,\infty)])$ 
for our case of $N=3$.
Next, by taking $L^{\infty}$-norms in (\ref{eq:H1}), we derive the inequality 
\begin{eqnarray}
||\mathcal{T}[C_0]||_{\infty}\leq 1 + \chi (1+\varrho)^{(1+2\sigma)}||A||_{\infty}.
\label{eq:H2}
\end{eqnarray}
From (\ref{eq:H2}), under the assumption
\begin{eqnarray}
\chi \leq \frac{\varrho}{(1+\varrho)^{(1+2\sigma)}|| A(t)||_{\infty}}:=\chi_1,
\label{eq:H3}
\end{eqnarray}
it follows that, 
$$||\mathcal{T}[C_0]||_{\infty} \leq 1+\varrho.$$
Thus, under restriction (\ref{eq:H3}) on $\chi$, we have proved the claim $\mathcal{T} : B \to B$.

We proceed by showing conditions on $\chi$ under which the map $\mathcal{T} : B \to B$ is a contraction.  First, note that the power nonlinearity (\ref{nonl1}) is of the form $F(z)=g(|z|^2)z$ with $g(r)=r^{\sigma}$, $\sigma>0$.  
Next, we recall that for any $F:\mathbb{C}\rightarrow\mathbb{C}$ which
takes the form $F(z)=g(|z|^2)z$, with $g$ real and
sufficiently smooth, the following relation holds
\begin{eqnarray}
\label{eq9}
&&F(\zeta)-F(\xi)=\int_{0}^{1}(\zeta-\xi)(g(r)+rg'(r))d\theta\nonumber\\
&&+\int_{0}^{1}(\overline{\zeta}-\overline{\xi})\Phi^2 g'(r)d\theta,
\end{eqnarray}
for any $\zeta,\;\xi\in \mathbb{C}$,where $\Phi=\theta \zeta
+(1-\theta)\xi$, $\theta\in (0,1)$ and $r=|\Phi|^2$ (see \cite[pg.
202]{GiVel96}). Here, $\overline{\zeta}$ denotes the complex conjugate of $\zeta\in\mathbb{C}$.

Let $C_0(t)$ and $Q_0(t)$ be two elements of $B$. Then, from (\ref{eq8}) we observe that
\begin{eqnarray}
\label{eq10}
&&\mathcal{T}[C_0(t)]-\mathcal{T}[Q_0(t)]\nonumber\\
&&=\mathrm{i}\chi\int_{0}^{t}J_0[2V(t-s)]^3\left[F(C_0(s))-F(Q_0(s))\right]ds.\;\;\;\;\;\;\;\;\;
\end{eqnarray}
Applying (\ref{eq9}) for $\zeta=C_0(t)$, $\xi=Q_0(t)$ one finds that
\begin{eqnarray}
\label{eq11}
&&F(C_0(t))-F(Q_0(t))\nonumber\\
&&=\int_0^1[(\sigma
+1)(C_0(t)-Q_0(t))|\Phi|^{2\sigma}d\theta\nonumber\\
&&+\sigma\int_0^1(\overline{C_0(t)}-\overline{Q_0(t)})\Phi^2|\Phi|^{2\sigma
-2}]d\theta.
\end{eqnarray}
Since
$||\Phi||_{\infty}\leq 1$, we get from (\ref{eq11}), the inequality
\begin{eqnarray}
\label{eq12}
&&|F(C_0(t))-F(Q_0(t))| \nonumber\\
&&\leq (2\sigma+1)\int_0^1|\Phi|^{2\sigma}|C_0(t)-Q_0(t)|d\theta\nonumber\\
&\leq&
(2\sigma+1)
\int_0^1||\Phi||_{\infty}^{2\sigma}|C_0(t)-Q_0(t)|d\theta\nonumber\\
&\leq&
(1+\varrho)^{2\sigma}(2\sigma+1)|C_0(t)-Q_0(t)|.
\end{eqnarray}
Inserting (\ref{eq12}) into (\ref{eq10}) and taking $L^{\infty}$-norms, we observe that
\begin{eqnarray}
\label{eq13}
&&||\mathcal{T}[C_0]-\mathcal{T}[Q_0]||_{\infty}\nonumber\\
&&=\mathrm{ess\;sup}_{t\in[0,\infty)}|\mathrm{i}\chi\int_{0}^{t}J_0[2V(t-s)]^3\nonumber\\
&&\times\left[F(C_0(s))-F(Q_0(s))\right]ds|\nonumber\\
&\leq& ||A||_{\infty}(1+\varrho)^{2\sigma}(2\sigma+1)\chi||C_0-Q_0||_{\infty},
\end{eqnarray}
It is clear from (\ref{eq13}) that the map $\mathcal{T}$ is a contraction if 
\begin{eqnarray}
\label{eq14}
\chi<\frac{1}{||A||_{\infty}(1+\varrho)^{2\sigma}(2\sigma+1)}:=\chi_2.
\end{eqnarray}
We may consider the value $\chi_{\mathrm{crit}}:=\min\{\chi_1,\chi_2\}$, where $\chi_1,\chi_2$ are defined in (\ref{eq:H3}) and (\ref{eq14}), respectively.  Some calculus around $\varrho$, implies that
\begin{eqnarray}
\label{eq14s}
\frac{1}{[(1+2\sigma)(1+\frac{1}{2\sigma})^{2\sigma}||A||_{\infty}]}=\chi_{\mathrm{crit}}.
\end{eqnarray}
Thus, we have proved the existence of $\chi_{\sigma}$ such that, if $\chi<\chi_{\sigma}$, the map satisfies $\mathcal{T} : B \to B$ and is a contraction, e.g., both conditions of the contraction mapping are satisfied. Thus, $\mathcal{T}$ has a unique fixed point $C_0\in B$, with the constant $\varrho$ in $B$ determined as $\varrho=\frac{1}{2\sigma}$.

We conclude the proof, by showing that the fixed point $C_0$  satisfies 
\begin{eqnarray}
\label{NKPGK0}
\lim_{t\to\infty}|C_0(t)|=0.
\end{eqnarray}
 The dynamics over an infinite 
time horizon $t \rightarrow \infty$ can be thought in a number of
ways. One such is to consider the map $\mathcal{T}[C_0(t)]$ with $t \rightarrow
\infty$. In this case, we observe that for an arbitrary $X_0\in B$,
\begin{eqnarray}
\label{NKPGK1}
&&\lim_{t\rightarrow\infty}\mathcal{T}[X_0(t)]\nonumber\\
&=&\lim_{t\rightarrow\infty}\mathrm{i}\chi\int_{0}^{t}J_0[2V(t-s)]^3|X_0(s)|^{2\sigma}X_0(s)ds\;\;\;\;\;\;\nonumber\\
&=&\lim_{t\rightarrow\infty}\mathcal{T}_{\star}[X_0(t)].
\end{eqnarray}
We justify first that the right-hand side of (\ref{NKPGK1}) makes sense for any $X_0\in B$: note that the map $\mathcal{T}_{\star}$, given by the integral in (\ref{NKPGK0}),
\begin{eqnarray*}
\mathcal{T}_{\star}[u_0(t)]=\mathrm{i}\chi\int_{0}^{t}J_0[2V(t-s)]^3|u_0(s)|^{2\sigma}u_0(s)ds,
\end{eqnarray*} 
is well defined as a map $\mathcal{T}_{\star}:L^{\infty}([0, \infty))\rightarrow L^{\infty}([0,\infty))$for arbitrary large $t\in [0,\infty)$.  Indeed, for any $u_0\in L^{\infty}([0, T))$,  
\begin{eqnarray*}
||\mathcal{T}_{\star}[u_0]||_{\infty}&\leq& ||u_0||_{\infty}^{2\sigma+1}\int_{0}^t|J_0[2V(t-s)]|^3ds\\
&\leq& ||u_0||_{\infty}^{2\sigma+1}||A||_{\infty}<\infty.
\end{eqnarray*}
Furthermore, treating the map $\mathcal{T}_{\star}$ exactly as the map $\mathcal{T}$, we may prove that if $\chi<\chi_{\mathrm{crit}}$, where $\chi_{\mathrm{crit}}$ is the same given in (\ref{eq14s}), then $\mathcal{T}_{\star}:B\rightarrow B$,  e.g., $\mathcal{T}_{\star}[X_0]\in B$. Also if $\chi<\chi_{\mathrm{crit}}$, then $\mathcal{T}_{\star}:B\rightarrow B$, is a contraction, having a unique fixed point ${U}_0\in B$. However, since $\mathcal{T}_{\star}[0]=0$, by the uniqueness of the fixed point, we have $U_0=0$. 

Now, from the consequences of the contraction mapping theorem, for the $m$-iteration of $\mathcal{T}_{\star}:B\rightarrow B$, we have the following: for any $X_0\in B$,
and $m\in\mathbb{N}$, $\mathcal{T}_{\star}^{m}[X_0(t)]\in B$ and
\begin{eqnarray}
\label{NKPGK3}
\lim_{m\to \infty}\mathcal{T}_{\star}^{m}[X_0(t)]=U_0=0
\end{eqnarray}
Of course, the original map $\mathcal{T}:B\rightarrow B$, shares the same consequence but for its fixed point $C_0$: For any $X_0\in B$, $\mathcal{T}^{m}[X_0]\in B$, and since $C_0$ is its unique fixed point, we have 
\begin{eqnarray}
\label{NKPGK5}
\lim_{m\to \infty}\mathcal{T}^{m}[X_0(t)]=C_0(t).
\end{eqnarray}
Besides,  from (\ref{NKPGK1}), the maps $\mathcal{T}:B\to B$ and $\mathcal{T}_{\star}:B\to B$ are asymptotically the same as $t\rightarrow\infty$. Thus, (\ref{NKPGK1}) is also valid for their $m+1$-iterations, e.g.,
\begin{eqnarray*}
\lim_{t\rightarrow\infty}\mathcal{T}^{m+1}[X_0(t)]
=\lim_{t\rightarrow\infty}\mathcal{T}^{m+1}_{\star}[X_0(t)],
\end{eqnarray*}
and any $X_0\in B$. The last equality can be rewritten as
\begin{eqnarray}
\label{NKPGK6}
\lim_{t\rightarrow\infty}\mathcal{T}[\mathcal{T}^m[X_0(t)]]
=\lim_{t\rightarrow\infty}\mathcal{T}_{\star}[\mathcal{T}^m_{\star}[X_0(t)]],
\end{eqnarray}
for any $X_0\in B$. Now we may pass to the limit in (\ref{NKPGK6}): by using (\ref{NKPGK3}) and (\ref{NKPGK5}), we eventually get: 
\begin{eqnarray*}
\lim_{t\rightarrow\infty}C_0(t)&=&\lim_{t\rightarrow\infty}\mathcal{T}[C_0(t)]\\
&=&\lim_{t\rightarrow\infty}\mathcal{T}[\lim_{m\to \infty}\mathcal{T}^{m}[X_0(t)]]\nonumber\\
&=&\lim_{t\rightarrow\infty}\mathcal{T}_{\star}[\lim_{m\to \infty}\mathcal{T}^{m}_{\star}[X_0(t)]]\nonumber\\
&=&\lim_{t\rightarrow\infty}\mathcal{T}_{\star}[U_0(t)]]=0.
\end{eqnarray*}
Thus we have concluded that if $\chi<\chi_{\mathrm{crit}}$, the asymptotic behavior of the unique fixed point $C_0$ of the map $\mathcal{T}:B\rightarrow B$ is (\ref{NKPGK0}).

\emph{B. Saturable nonlinearity.} Applying (\ref{eq9}) in the case of the saturable nonlinearity (\ref{nonl2}) where $g(r)=\frac{r}{1+r}$, 
we have that
\begin{eqnarray*}
&&F(C_0(t))-F(Q_0(t))=
\int_0^1(C_{0} - Q_{0}) {|\Phi|^2 (2 + |\Phi|^2)\over{(1 + |\Phi|^2)^2}}d\theta\nonumber\\
&+&\int_0^1(\overline{C_{0}}-\overline{Q_{0}}) {|\Phi|^2\over{(1 + |\Phi|^2)^2}}d\theta, 
\end{eqnarray*}


from which the inequality
\begin{eqnarray}
\label{eq16}
&&|F(C_0(t))-F(Q_0(t))|\nonumber\\
&\leq& 
\int_0^1 (C_{0}-Q_{0}) {|\Phi|^2 (3 + |\Phi|^2)\over{(1+|\Phi|^2)^2}}d\theta\nonumber\\
&\leq& {9\over{8}}|C_0(t)-Q_0(t)|,
\end{eqnarray}
follows.  Then from (\ref{eq10}) and (\ref{eq16}) we get that
\begin{eqnarray}
\label{eq17}
||\mathcal{T}[C_0]-\mathcal{T}[Q_0]||_{\infty}
\leq \frac{9}{8}||A||_{\infty}\chi||C_0-Q_0||_{\infty}.
\end{eqnarray}
Hence, in the case of saturable nonlinearity,  the map $\mathcal{T}:B\rightarrow B$ will be a contraction if 
\begin{eqnarray}
\label{eq18}
\chi<\frac{8}{9||A||_{\infty}}=\chi_{\mathrm{crit}}.
\end{eqnarray}
To conclude that the unique fixed point satisfies $\lim_{t\to\infty}|C_0(t)|=0$,
we follow the same lines as in proof of case A. 
\ \ $\Box$

A comparison of (\ref{eq14s}) and (\ref{eq18}), shows that in the $3$D-lattice, the critical value for the power nonlinearity is less than the one for the saturable nonlinearity. This is an indication, that at least in some cases of $\sigma$, we may expect that the true critical points $\chi_c$  may share such an ordering. 

Let us also remark that Theorem \ref{th1},  does not imply directly that $\langle P\rangle=0$, unless a particular knowledge of the rate of decay of $|C_0(t)|^2$ is known. The possible relation of the diagnostics (\ref{eq0}) and (\ref{eq0a}) will be discussed in section \ref{SECV}.

Finally, it is evident from the proof of Theorem \ref{th1}, that the technical restriction on the dimension $N=3$ is imposed by the divergence of the relevant function $A(t)=\int_{0}^{t}\left|J_0[2V(t-s)]\right|^Nds$, when the dimensions $N=1,2$ are considered. 

{\em Approximating quantifications of $\chi_{\mathrm{crit}}$ for $N=3$.} 
\label{parA}
A common threshold  on $\chi$ can be derived for both nonlinearities, if we assume that for large times the flow governing $C_0(t)$ is asymptotically linear. 
In that case, more specifically, we assume that for 
large times $|C_0(t)|^2\cong \epsilon^2$, and $\epsilon>0$ is sufficiently small. Thus for the nonlinearities $F(z)=g(|z|^2)z$ the map (\ref{eq8}) is linearly approximated by the map
\begin{eqnarray}
\label{eq8lin}
\mathcal{T}[C_0(t)]\approx\mathrm{i}\chi g(\epsilon^2)\int_{0}^{t}J_0[2V(t-s)]^3C_0(s)ds,\nonumber\\ 
\end{eqnarray}
and for (\ref{eq8lin}), the analogue of (\ref{eq13}) is 
\begin{eqnarray}
\label{eq8alin}
||\mathcal{T}[C_0]-\mathcal{T}[Q_0]||_{\infty}
\leq ||A||_{\infty}g(\epsilon^2)\chi||C_0-Q_0||_{\infty}.\;\;\;\;\;\;
\end{eqnarray}
Note that for both types of nonlinearities, power and saturable, $0<g(\epsilon^2)<1$, hence
\begin{eqnarray*}
\label{eq13lin}
||\mathcal{T}[C_0]-\mathcal{T}[Q_0]||_{\infty}
\leq ||A||_{\infty}\chi||C_0-Q_0||_{\infty}.
\end{eqnarray*}
Then, the linear map (\ref{eq8lin}) is a contraction for both types of nonlinearities, if 
\begin{eqnarray}
\label{eq14lin}
\chi<\chi_{\mathrm{crit}}=\frac{1}{||A||_{\infty}}.
\end{eqnarray}
Note that the common value $\chi_{\mathrm{crit}}$ is greater than (\ref{eq18}) derived for the saturable nonlinearity. This was expected, since the saturable nonlinearity is sublinear.

It is also interesting that the asymptotically linear approximation may reveal different monotonicity properties of the threshold value for $\chi$ as a function of $\sigma$ in the case of the power nonlinearity. For instance, note from (\ref{eq14s}) that if $\sigma_1\geq\sigma_2$, then $\chi_{\mathrm{crit}}(\sigma_1)\leq\chi_{\mathrm{crit}}(\sigma_2)$, showing that the threshold $\chi_{\mathrm{crit}}$ derived from the contraction mapping theorem is a decreasing function of $\sigma$. 
In view of the contraction mapping theorem, this monotonicity is naturally expected, since it is associated with the smallness condition for the convergence of the iteration scheme associated to the integral equation (\ref{eq8}). This can be highlighted by recalling the approximating iteration scheme for (\ref{eq8})
\begin{eqnarray*}
C_0^{(n+1)}(t)&=&J_0(2Vt)^3\\
&+&\mathrm{i}\chi\int_0^tJ_0[2V(t-s)]^3|C_0^{(n)}(s)|^{2\sigma}C_0^{(n)}(s)ds,
\end{eqnarray*}
satisfying for the convergence to the unique fixed point $C_0(t)$, the error estimate
\begin{eqnarray*}
||C_0^{(n)}(t)||_{\infty}&\leq&\frac{K^n}{1-K}||C_0^{(1)}(t)-C_0^{(0)}(t)||_{\infty},\\ K(\chi,\sigma)&=&(1+2\sigma)(1+\frac{1}{2\sigma})^{2\sigma}||A||_{\infty}\chi.
\end{eqnarray*}
Increasing the strength of the nonlinearity as $\sigma$ is increasing, the ansatz of $K(\chi,\sigma)$ justifies that $\chi$ should be decreasing so that $K(\chi, \sigma)<1$, to guarantee convergence of the iteration scheme.
This requirement for convergence, is giving rise, in turn, to the critical value $\chi_{\mathrm{crit}}$ of (\ref{eq14s}).

However, from the approximate argument of (\ref{eq8alin}), it follows that the linearized map is a contraction if
\begin{eqnarray}
\label{monp1}
\chi<\frac{1}{||A||_{\infty}\epsilon^{2\sigma}}=\hat{\chi}(\sigma),\;\;0<\epsilon<1.
\end{eqnarray}
Since $1/\epsilon>1$, the threshold for the linearized map satisfies the monotonicity property
\begin{eqnarray}
\label{monp2}
\hat{\chi}(\sigma_1)>\hat{\chi}(\sigma_2),\;\;\mbox{for $\sigma_1>\sigma_2$.}
\end{eqnarray}
Thus, when  the decay of $C_0$ is slow in amplitude, we expect that for the power nonlinearity, the true threshold $\chi_c$ will be an increasing function of $\sigma$. 

The discussion above,  between the difference of the monotonicity properties of the critical value $\chi_{\mathrm{crit}}$ derived by the contraction mapping argument of Theorem \ref{th1}, and the  threshold $\hat{\chi}$ for the linear approximation, indicates on the one side, that $\chi_{\mathrm{crit}}$ can serve as a lower bound for the true threshold but cannot be sharp. On the other side, the lack of sharpness of $\chi_{\mathrm{crit}}$ is justified by the fact that a violation of the condition of Theorem \ref{th1}, that is, assuming $\chi>\chi_{\mathrm{crit}}$, does {\it not} imply the non-existence of another nontrivial fixed point $C_0$ 
 and that $||C_0||_{\infty}\neq 0$, for large times.  In other words, although the smallness conditions on $\chi$ stemming from Theorem \ref{th1} are sufficient for establishing 
that $|C_0(t)|$ tends to $0$ for large $t$, they are not necessary. 

On account of the common threshold (\ref{eq14lin}) for both types of nonlinearities, we may summarize on the following lower bound for the true critical value,
\begin{eqnarray}
\label{LOW}
\frac{1}{||A||_{\infty}}\leq \chi_{c}.
\end{eqnarray}

\section{Minimal values on $\chi$ for the formation of stationary states}
\label{SECIV}
\setcounter{equation}{0}
\subsection{Minimal values for the creation of stationary states from the dynamical equation}
A first attempt to approximate  $\chi_{c}$ closer to its real value is provided 
by noting that, according to numerical results, when $\chi>\chi_{c}$,  it seems that $|C _{0}(t)|$ is approaching  a constant value. That is, $C_{0}(t)$ resembles a ``stationary'' state (i.e., a standing wave in the context of
the DNLS). 

\paragraph{Lattice Dimension $N=1$.} 
\label{dynN1}
We shall implement such an assumption first in the $1$D-lattice. 
Thus, after a long time, we assume that in the trapping regime $C_0(t)\approx \alpha \exp(i\beta t)$. For example, replacing the stationary state in the long-time approximation for the cubic nonlinearity
\begin{eqnarray}
\label{MarioP}
C_{0}(t)\approx i \chi \int_{0}^{t} J_{0}[2 V (t-s)] |C_{0}(s)|^2 C_{0}(s) ds,
\end{eqnarray}
and after splitting into real and imaginary parts, one obtains:
\begin{eqnarray*}
1 &=&  \chi |\alpha|^2 \int_{0}^{\infty} J_{0}(2 V z) \sin(\beta z) dz,\\
0& =& \int_{0}^{\infty} J_{0}(2 V z) \cos(\beta z) dz.
\end{eqnarray*}
Using basic Bessel integral properties \cite[Sec. 6.671, eq. (7)]{GR} one concludes $2 V < \beta$ and
\begin{eqnarray}
\label{MarioP1}
 1 = { \chi|\alpha|^2\over{\sqrt{\beta^2-(2 V)^2}}}.
\end{eqnarray}
This implies,
\begin{eqnarray*}
\chi = {\sqrt{\beta^2 - (2 V)^2}\over{|\alpha|^2}}.
\end{eqnarray*}
In particular, when $V\rightarrow 0$, there is complete self-trapping and $|\alpha|\rightarrow 1$ 
implying $\chi\rightarrow \beta > 2V$. Therefore, inside the trapping regime, $\chi > 2 V$ at long times. The critical value $2V$ also coincides with the minimum value to create a nonlinear stationary state,  and  the lower bound 
\begin{eqnarray}
\label{INT}
\chi_{\mathrm{stationary}}=2V<\chi_{c},
\end{eqnarray} 
accordingly becomes meaningful. Intuitively, the critical value $\chi_{\mathrm{stationary}}$ for the formation of the stationary mode, should always be smaller than the corresponding value $\chi_{c}$ for the dynamical problem, since the former value stems from a variational principle. 

A similar argument, can be applied in the case of the saturable nonlinearity (\ref{nonl2}). Indeed, in this case, the counterparts of (\ref{MarioP}) and (\ref{MarioP1}) are
\begin{eqnarray}
\label{MarioS}
C_{0}(t)\approx i \chi \int_{0}^{t} J_{0}[2 V (t-s)]  \frac{|C_{0}(s)|^2 C_{0}(s)}{1+|C_{0}(s)|^2} ds
\end{eqnarray}
and 
\begin{eqnarray}
\label{MarioS1}
 1 = \frac{\chi |\alpha|^2}{1+|\alpha|^2}\frac{1}{\sqrt{\beta^2-(2 V)^2}},
\end{eqnarray}
for $\beta>2V$. Then, (\ref{MarioS1}) implies that
\begin{eqnarray*}
\chi=\sqrt{\beta^2-(2 V)^2}{(1+|\alpha|^2)\over{|\alpha|^2}}.
\end{eqnarray*}
Accordingly, when $V\rightarrow 0$, we have $|\alpha|\rightarrow 1$ in the complete self-trapping, and $\chi\rightarrow 2\beta>4V$. Hence, in the case of the saturable nonlinearity we have the lower bound
\begin{eqnarray}
\label{INTS}
\chi_{\mathrm{s, stationary}}=4V<\chi_{c}.
\end{eqnarray} 
\paragraph{Lattice dimension $N=2$.} 
\label{dynN2b}
For the extension of the above argument to the $2$D-lattice, the point of departure is the $2$D-counterpart of eq.~(\ref{MarioP})
\begin{eqnarray}
\label{N22}
C_{0}(t)\approx i \chi \int_{0}^{t} J_{0}[2 V (t-s)]^2 |C_{0}(s)|^2 C_{0}(s) ds.
\end{eqnarray}
Splitting (\ref{N22}) into real and imaginary parts, one obtains:
\begin{eqnarray}
\label{N23}
1 &=&  \chi |\alpha|^2 \int_{0}^{\infty} J_{0}(2 V z)^2 \sin(\beta z) dz,\\
\label{N24}
0& =& \int_{0}^{\infty} J_{0}(2 V z)^2 \cos(\beta z) dz.
\end{eqnarray}
With the change of variables $z=x/2V$, (\ref{N23}) becomes
\begin{eqnarray}
\label{NK25}
1 &=&  \frac{\chi |\alpha|^2}{2V} \int_{0}^{\infty} J_{0}(x)^2 \sin\left(\frac{\beta x}{2V}\right)dx.
\end{eqnarray}
Thus, the integral in the right-hand side of (\ref{NK25}) is of the form  (\ref{INK}) of \ref{AppPb}, with $A=\frac{\beta}{4V}$. Applying the branch of (\ref{N32}) for $A>1$ to this integral, we find that
\begin{eqnarray}
\label{NK26}
1=\frac{2\chi|\alpha|^2}{\pi\beta}\int^{\frac{\pi}{2}}_0\frac{d\phi}
{\sqrt{1-\frac{(4V)^2}{\beta^2}\sin^2\phi}},\;\;\mbox{for $\beta>4V$}.\;\;\;\;\;\;\;
\end{eqnarray}
When $V\rightarrow 0$, we have $|\alpha|\rightarrow 1$ in the complete self-trapping, and 
\begin{eqnarray}
\label{NK27}
\int^{\frac{\pi}{2}}_0\frac{d\phi}
{\sqrt{1-\frac{(4V)^2}{\beta^2}\sin^2\phi}}\rightarrow \frac{\pi}{2}.
\end{eqnarray}
In this case, (\ref{NK26}) and (\ref{NK27}) imply that $\chi\rightarrow \beta>4V$. Therefore, when $N=2$, for the cubic nonlinearity 
\begin{eqnarray}
\label{NK28}
\chi_{\mathrm{stationary}}=4V<\chi_{c}.
\end{eqnarray} 
When the saturable nonlinearity (\ref{nonl2}) is considered, the counterparts of (\ref{N22}) and (\ref{NK26}) are
\begin{eqnarray*}
C_{0}(t)\approx i \chi \int_{0}^{t} J_{0}[2 V (t-s)]^2 |C_{0}(s)|^2 C_{0}(s) ds,
\end{eqnarray*}
and 
\begin{eqnarray*}
1=\frac{2\chi|\alpha|^2}{\pi\beta(1+|\alpha|^2)}\int^{\frac{\pi}{2}}_0\frac{d\phi}
{\sqrt{1-\frac{(4V)^2}{\beta^2}\sin^2\phi}},\;\;\mbox{for $\beta>4V$}.\;\;\;\;\;\;\;
\end{eqnarray*}
In the complete self-trapping where $|\alpha|\rightarrow 1$ when $V\rightarrow 0$, the latter implies that $\chi\rightarrow 2\beta>8V$. Thus in the $2$D-lattice, for the saturable nonlinearity we have the lower bound
\begin{eqnarray}
\label{NK31}
\chi_{\mathrm{s,stationary}}=8V<\chi_{c}.
\end{eqnarray}  
A comparison of (\ref{INT}) with (\ref{NK28}) for the power nonlinearity, and (\ref{INTS}) with (\ref{NK31}) may suggest that in the full discrete regime of moderate values of $V$, we may have the lower bounds
\begin{eqnarray*}
\chi_{\mathrm{stationary}}&=&2NV<\chi_{c},\\
\chi_{\mathrm{s,stationary}}&=&4NV<\chi_{c}.
\end{eqnarray*}
for the power and the saturable nonlinearities in the $N$D-lattices. We will verify that this is the case, in the next paragraph.
\subsection{Variational methods for the full stationary problem}
Motivated by the above observations, that the lower bounds (\ref{INT})-(\ref{INTS}) are stemming by variational principles, we will derive critical values on $\chi$ for the formation of stationary states 
\begin{equation}
\label{TP1}
C_n(t)=\mathrm{e}^{\mathrm{i}\beta t}\Phi_n,\;\;\beta\in\mathbb{R},\;\;\Phi_n\in\mathbb{C},
\end{equation}
in higher dimensional lattices, by  implementing variational methods.
Such critical values will be derived by the consideration of the full stationary problem associated with the DNLS equation (\ref{eq:1}).  Here,  $n\in\mathbb{Z}^N$ (see the complementary section \ref{AppPa} for various notations on the $N$-dimensional lattice $\mathbb{Z}^N$).
We consider the DNLS equation (\ref{eq:1}) in the case of site-independent coupling $V_{n,m}=V$. Note first, that the change of variables $C_n(t)\rightarrow\mathrm{e}^{-2\mathrm{i} N t}C_n(t)$, the application of the staggering transformation $C_n\rightarrow (-1)^{|n|}C_n,\;\;\; |n|=\sum_{i=1}^N n_i$, and the  $t \rightarrow -t$ transformation
(taking advantage of the problem's time
reversibility), bring (\ref{eq:1}) to the form
\begin{eqnarray}
\label{DNLSd}
\mathrm{i}\dot{C}_n&-&V\Delta_d C_n-\chi\delta_{n,0}F(C_n)=0,\;\;||n||\leq K,\\
C_n&=&0,\;\;||n||>K,\nonumber
\end{eqnarray}
where  
$\Delta_d\psi_n$ stands for the $N$-dimensional discrete Laplacian
\begin{eqnarray}
\label{DiscLap}
\Delta_d\psi_{n\in\mathbb{Z}^N}=\sum_{m\in \mathcal{N}_n}\psi_m-2N\psi_n.
\end{eqnarray}
In (\ref{DiscLap}), $\mathcal{N}_n$ denotes the set of $2N$ nearest neighbors of
the point in $\mathbb{Z}^N$ with label $n$. Also $||n||=\max_{1\leq i\leq N}|n_i|$. Note that in (\ref{DNLSd}) we have considered Dirichlet boundary conditions, however, this is not a loss of generality for our purposes, due to reasons that will be explained below.

Next, substitution of the solution (\ref{TP1}) to (\ref{DNLSd}) shows that $\Phi_n$ satisfies the stationary problem 
\begin{eqnarray}
\label{DNLSd1}
-\beta\Phi_n&-&V\Delta_d \Phi_n-\chi\delta_{n,0}F(\Phi_n)=0,\;\;||n||\leq K,\;\;\;\;\;\;\;\\
\Phi_n&=&0,\;\;||n||>K.\nonumber
\end{eqnarray}
Since (\ref{eq5a}) holds, we are interested in stationary states (\ref{TP1}) of unit energy, e.g., 
\begin{eqnarray}
\label{TP2}
\sum_{||n||\leq K}|\Phi_n|^2=1. 
\end{eqnarray}
In what follows, $\lambda_1$, stands for the first (principal) eigenvalue of the discrete Laplacian (see eq. (\ref{DLap}, section \ref{AppPa}). For the formation of such states a critical value on $\chi$ exists, as stated in the following 
\begin{theorem}
\label{Thstates}
Nontrivial stationary states (\ref{TP1}) with energy (\ref{TP2}), exist when\newline
A. (Power nonlinearity):\begin{eqnarray}
\label{Crit_chi_P}
\lambda_1-
\beta\leq \chi,\;\;\left\{
  \begin{array}{ll}
    \beta<\lambda_1,\;\hbox{if $\beta>0$,}\\
    \forall\beta<0.
  \end{array}
\right.\;\;\;\;\;
\end{eqnarray} 
\newline
B. (Saturable nonlinearity): 
\begin{eqnarray}
\label{Crit_chi_s}
4VN<\chi,\;\;\;\forall\beta<0.
\end{eqnarray} 
\end{theorem}
\textbf{Proof:} \emph{A. Power nonlinearity.} We consider the  variational problem on $\ell^2(\mathbb{Z}^N_K)$,
\begin{eqnarray}
\label{infsigmaE}
\inf\left\{\mathcal{H}[\Phi]\;:\sum_{||n||\leq K}|\Phi_n|^{2}=
  1\right\},
\end{eqnarray}
where 
\begin{eqnarray*}
\label{Hamsigma}
\mathcal{H}[\Phi]&=&(-V\Delta_d\Phi,\Phi)_2-\frac{\chi}{\sigma+1}
\sum_{||n||\leq K}\delta_{n,0}|\Phi_n|^{2\sigma+2}\nonumber\\
&=&(-V\Delta_d\Phi,\Phi)_2-\frac{\chi}{\sigma+1}|\Phi_0|^{2\sigma+2},
\end{eqnarray*}

denotes the Hamiltonian associated with the DNLS equation (\ref{DNLSd}). Let also 
$$B=\left\{\Phi\in\ell^2(\mathbb{Z}^N_K)\;:\;
  \sum_{||n||\leq K}|\Phi_n|^{2}=R^2\right\}.$$
Working along the lines of \cite{p1}-\cite{p4}, it can be shown that $\mathcal{H}:B\rightarrow\mathbb{R}$ is  a $C^1$-functional which is bounded from below, and that $\mathcal{H}$ attains its infimum at a point
$\hat{\Phi}$ in $B$, the solution of the variational problem (\ref{infsigmaE}). Along the same lines, it can be shown that the Lagrange multiplier rule is applicable (see \cite[Theorem 2.1]{p3} or \cite[Theorem 2.3]{p1}), justifying the existence of a parameter 
$\beta\in\mathbb{R}$, such that 
\begin{eqnarray}
\label{Esigma}
&&(-V\Delta_d\hat{\Phi},\Psi)_{2}-\chi|\hat{\Phi}_0|^{2\sigma}\hat{\Phi}_0
\overline{\Psi_0}\\
&&-\beta\mathrm{Re}
\sum_{||n||\leq K}\hat{ \Phi}_n\overline{\Psi}_n=0\nonumber,\;\;
\mbox{for all}\;\;\Psi\in\ell^2(\mathbb{Z}^N_K).
\end{eqnarray}
Equation (\ref{Esigma}) shows that $\Phi$ satisfies the Euler-Lagrange equation (\ref{DNLSd1}). Setting $\Psi=\hat{\Phi}$ in
(\ref{Esigma}), we find that
\begin{eqnarray}
\label{Esigma2}
(-V\Delta_d\hat{\Phi},\hat{\Phi})_{2}-
\beta=\chi|\hat{\Phi}_0|^{2\sigma+2},\;\;\;\;\;
\end{eqnarray}
since $\Phi$ has unit $l^2$ norm, as a solution of the variational problem (\ref{infsigmaE}).

A lot of useful information is included in (\ref{Esigma2}). First we note that since $\chi>0$, it holds that 
\begin{eqnarray}
\label{Esigma3}
(-V\Delta_d\hat{\Phi},\hat{\Phi})_{2}-
\beta\geq 0,
\end{eqnarray}
and $|\hat{\Phi}_0|^2$ satisfies
\begin{eqnarray}
\label{Esigma3a}
\left(\frac{(-V\Delta_d\hat{\Phi},\hat{\Phi})_{2}-
\beta}{\chi}\right)^{\frac{1}{\sigma+1}}=|\hat{\Phi}_0|^2.
\end{eqnarray}
Now, from (\ref{Esigma3a}) we can obtain estimates for $|\Phi_0|^2$, as well as for $\chi$ in terms of $\lambda_1$. Indeed, by using (\ref{crucequiv}) to estimate from below the right-hand side of (\ref{Esigma3}),
we get the lower bound for $|\hat{\Phi}_0|^2$
\begin{eqnarray}
\label{Esigma4}
\left(\frac{\lambda_1-
\beta}{\chi}\right)^{\frac{1}{\sigma+1}}\leq |\hat{\Phi}_0|^2,\;\;\left\{
  \begin{array}{ll}
    \beta<\lambda_1,\;\hbox{if $\beta>0$,}\\
    \forall\beta<0.
  \end{array}
\right.\;\;\;\;\;
\end{eqnarray} 
On the other hand, since the stationary state satisfies (\ref{TP2}), we have that $|\hat{\Phi}_0|^2\leq 1$, and (\ref{Esigma4}) implies that $\chi$ should satisfy 
\begin{eqnarray*}
\lambda_1-
\beta\leq \chi,\;\;\left\{
  \begin{array}{ll}
    \beta<\lambda_1,\;\hbox{if $\beta>0$,}\\
    \forall\beta<0,
  \end{array}
\right.\;\;\;\;\;
\end{eqnarray*} 
as claimed. Note also that from (\ref{Esigma2}) and (\ref{crucequiv}), 
\begin{eqnarray}
\label{Esigma6}
(\lambda_1-
\beta)|\hat{\Phi}_0|^{-2(\sigma+1)}\leq \chi,\;\;\left\{
  \begin{array}{ll}
    \beta<\lambda_1,\;\hbox{if $\beta>0$,}\\
    \forall\beta<0.
  \end{array}
\right.\;\;\;\;\;\;\;\;
\end{eqnarray} 
\newline
\emph{B. Saturable nonlinearity.} In the case of the saturable nonlinearity, the Hamiltonian is
\begin{eqnarray}
\label{HamsigmaS}
\mathcal{H}_s[\Phi]&=&(-V\Delta_d\Phi,\Phi)_2\nonumber\\
&-&\chi
\sum_{||n||\leq K}\delta_{n,0}(|\Phi|^2-\log(1+|\Phi_n|^2)\nonumber\\
&=&(-V\Delta_d\Phi,\Phi)_2\nonumber\\
&-&\chi(|\Phi_{0}|^2-\log(1+|\Phi_0|^2)).
\end{eqnarray} 
Following the same arguments as in case A (see also \cite[Theorem 3.3, pg. 456]{p4}), it can be shown that $\mathcal{H}_s$ attains its infimum at a point
$\hat{\Phi}$ in $B$, the solution of the variational problem (\ref{infsigmaE}), with the functional $\mathcal{H}$ replaced by $\mathcal{H}_s$. This time, the minimizer 
$\hat{\Phi}$ and the real parameter $\beta$ satisfy the equation
\begin{eqnarray}
\label{EsigmaS}
&&(-V\Delta_d\hat{\Phi},\Psi)_{2}-\chi\frac{\hat{\Phi}_0\overline{\Psi_0} |\Phi_{0}|^2}{1+|\hat{\Phi}_0|^2}
\\
&&-\beta\mathrm{Re}
\sum_{||n||\leq K}\hat{ \Phi}_n\overline{\Psi}_n=0\nonumber,\;\;
\mbox{for all}\;\;\Psi\in\ell^2(\mathbb{Z}^N_K).
\end{eqnarray}
Equation (\ref{EsigmaS}) shows that $\Phi$ satisfies the Euler-Lagrange equation (\ref{DNLSd1}) in the case of the saturable nonlinearity. Setting $\Psi=\hat{\Phi}$ in
(\ref{EsigmaS}), we find that
\begin{eqnarray}
\label{EsigmaS2}
(-V\Delta_d\hat{\Phi},\hat{\Phi})_{2}-
\chi\frac{|\hat{\Phi}_0|^{4}}
{1+|\hat{\Phi}_0|^{2}}=\beta.\;\;\;\;\;
\end{eqnarray}
By using the right-hand side of (\ref{crucequiv}), we derive the equation 
\begin{eqnarray}
\label{EsigmaS3}
(-V\Delta_d\hat{\Phi},\hat{\Phi})_{2}-
\chi\frac{|\hat{\Phi}_0|^{4}}
{1+|\hat{\Phi}_0|^{2}}\leq 4VN-\chi\frac{|\hat{\Phi}_0|^{4}}
{1+|\hat{\Phi}_0|^{2}}.\;\;\;\;\;\;\;\;\;\,
\end{eqnarray}
From (\ref{EsigmaS2}) and (\ref{EsigmaS3}) it follows that $\beta<0$, if the right-hand side of (\ref{EsigmaS3}) is negative, e.g., 
\begin{eqnarray}
\label{EsigmaS4}
\frac{4VN(1+|\hat{\Phi}_0|^{2})}{|\hat{\Phi}_0|^{4}}
<\chi.
\end{eqnarray}
It readily follows from  (\ref{EsigmaS4}) that $\chi$ satisfies
\begin{eqnarray*}
4VN<\chi,
\end{eqnarray*}
as claimed. Also, from (\ref{EsigmaS4}) we have the inequality
\begin{eqnarray*}
\label{SBound1}
\frac{4VN}{\chi} +  \frac{4VN}{\chi}|\hat{\Phi}_0|^2 < |\hat{\Phi}_0|^4.
\end{eqnarray*}
Inequality (\ref{SBound1}) implies that 
\begin{eqnarray}
\label{SBound2}
|\hat{\Phi}_0|^2 &>& \frac{2VN}{\chi} + 2\sqrt{\left(\frac{VN}{\chi}\right)^2+\frac{VN}{\chi}}
\nonumber\\
&>& \frac{4VN}{\chi},
\end{eqnarray}
which provides a lower bound for $|\hat{\Phi}_0|^2$ in the case of the saturable nonlinearity.\ \ \ $\Box$

Further quantifications on the critical value for $\chi$ for the formation of stationary states claimed in Theorem \ref{Thstates}, can be derived when the ``cut off''  energy approximation procedure of \cite{p1} for stationary states (\ref{TP1}) is applied. This procedure is analyzed in detail in \cite[Section 4.1.1, pg. 225-230]{p1} and approximates the contribution of the linear part to the energy of the stationary state, by taking into account its true localization length. 

We briefly discuss this procedure: we may extend \cite[Proposition 4.1 \& Remarks 4.2-4.4, pg. 227]{p1} on the $N$-dimensional unit cube $\mathcal{Q}$ (see section \ref{AppPa}), by assuming that the energy of the stationary state is mainly concentrated in $\mathcal{Q}$. This is certainly true for stationary states centered around the center of $\mathcal{Q}$. 
The result is to consider a variant of the stationary problem \ref{DNLSd1}, only for the sites included in $\mathcal{Q}$, supplemented with Dirichlet boundary conditions on the edges of the cube $\mathcal{Q}'$ consisting of the adjacent sites of $\mathcal{Q}$. For this restricted variant of the problem \ref{DNLSd1}, we repeat the proof of Theorem \ref{Thstates}. The difference with the full problem is that we estimate the contribution of the linear part to the energy of the stationary state, by the principal eigenvalue $\lambda_1$ of the discrete Laplacian (\ref{DLap}) on $\mathcal{Q}'$, as follows: 
\begin{itemize}
\item $\lambda_1= 4NV\sin^2\left(\frac{\pi}{4}\right)=2NV$ for $1\leq V\leq4$, (defined  here as the discrete regime),
\item $\lambda_1= 4NV\sin^2\left(\frac{\pi}{2\sqrt{V}}\right)$ for $V>4$ (defined here as a continuous regime).
\end{itemize}
It should be remarked that the above procedure is independent of the size of the lattice, since the cube $\mathcal{Q}'$ where the procedure takes place, is the same in either the finite or the infinite lattice. As an outcome, we have
\begin{corollary}
\label{ThstatesC}
Stationary states (\ref{TP1}) with arbitrary $\beta<0$ with energy (\ref{TP2}), exist when\newline
A. (Power nonlinearity):\begin{eqnarray*}
&&\chi_{\sigma,N,V}=2NV<\chi,\;\;1\leq V\leq 4\;\;\mbox{(discrete regime)},\\
&&\chi_{\sigma,N,V}=4NV\sin^2\left(\frac{\pi}{2\sqrt{V}}\right)<\chi,\;\;V>4\\
&&\mbox{(continuous regime)}.
\end{eqnarray*} 
\newline
B. (Saturable nonlinearity): $\chi_{s,N,V}=4VN<\chi$, for all $V>0$.
\end{corollary}
Theorem \ref{Thstates} and corollary \ref{Thstates} reveal some interesting quantitative properties of the 
minimum value for the formation of stationary states in the $N$-dimensional lattice. These properties are summarized in 
\begin{corollary}
\label{CV2}
\begin{enumerate}
\item The minimum values for the formation of stationary states are increasing functions of the dimension of the lattice $N$, in both cases of nonlinearities.
\item The minimum value in the case of the power nonlinearity is an increasing function of the nonlinearity exponent $\sigma$.
\end{enumerate}
\end{corollary}
\textbf{Proof:} Property 1 is an immediate consequence of Theorem \ref{Thstates} and Corollary \ref{ThstatesC}. Property 2, follows directly from the lower bound on $\chi$ given in (\ref{Esigma6}), since $|\Phi_0|<1$. \ \ $\Box$

The results of Theorem \ref{Thstates} 
are closely related to those derived by the application of local bifurcation theory; the conditions on $\chi$ for the formation of stationary states are close to those for their bifurcation from the principal eigenvalue  $\lambda_1$, \cite{ZeiFP,bif1}.  Additionally, the relevance of the conditions of \ref{Thstates} with the dynamical self-trapping is due to the fact that the conditions refer to the existence of stationary states as global minimizers of the Hamiltonian energy, i.e., ground states, suggesting their dynamical stability. Actually, these are the states which may bifurcate from the principal eigenvalue (see \cite{GV} for applications in NLS lattices in the context of BEC's).  

The qualitative and quantitative theoretical predictions proved and discussed in this section, will now be tested by numerical simulations. 
\section{Numerical Results}
\label{SECV}
\begin{figure}
\begin{center}
    \begin{tabular}{cc}
    \includegraphics[scale=0.34]{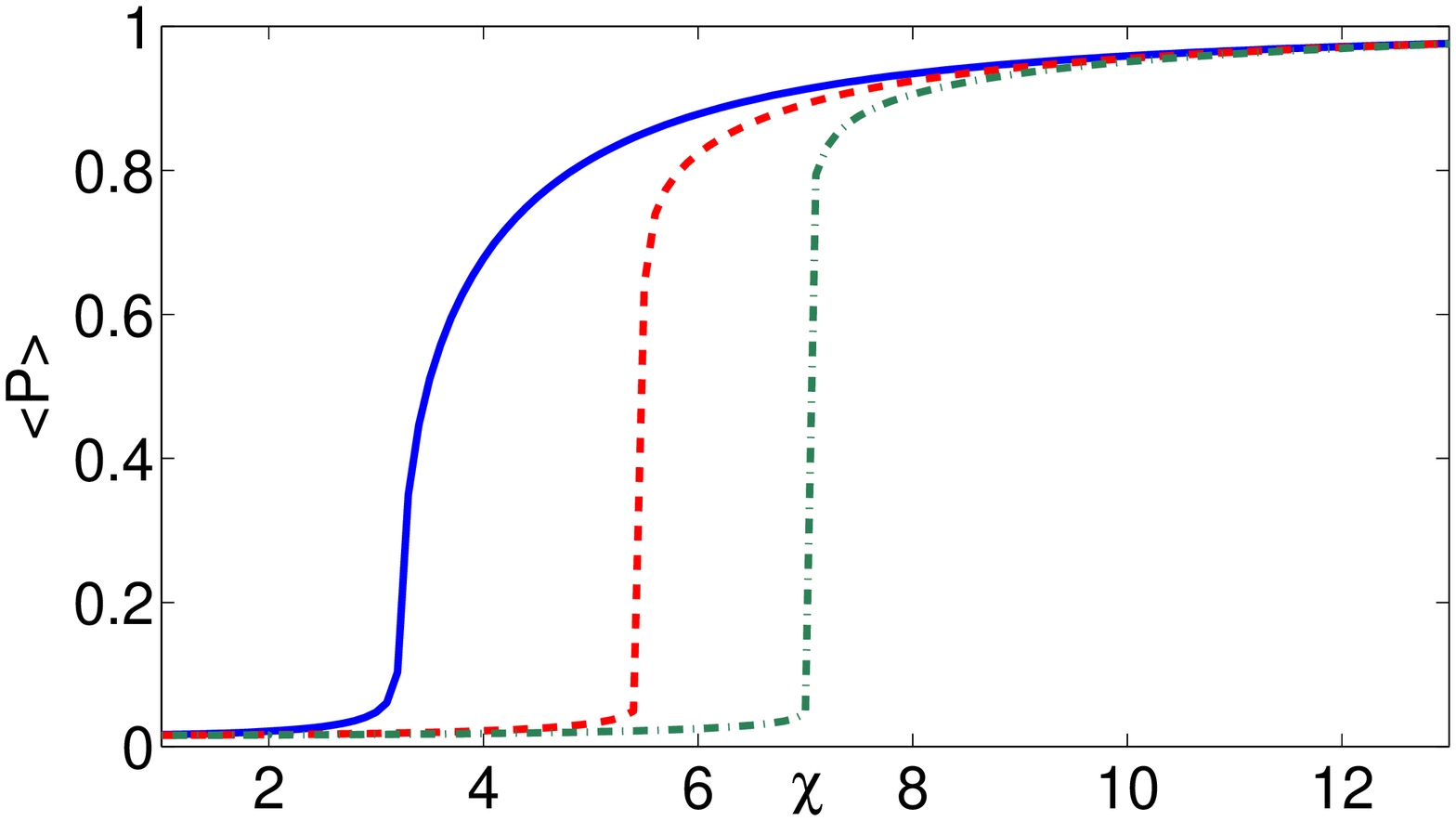}\\
    \includegraphics[scale=0.34]{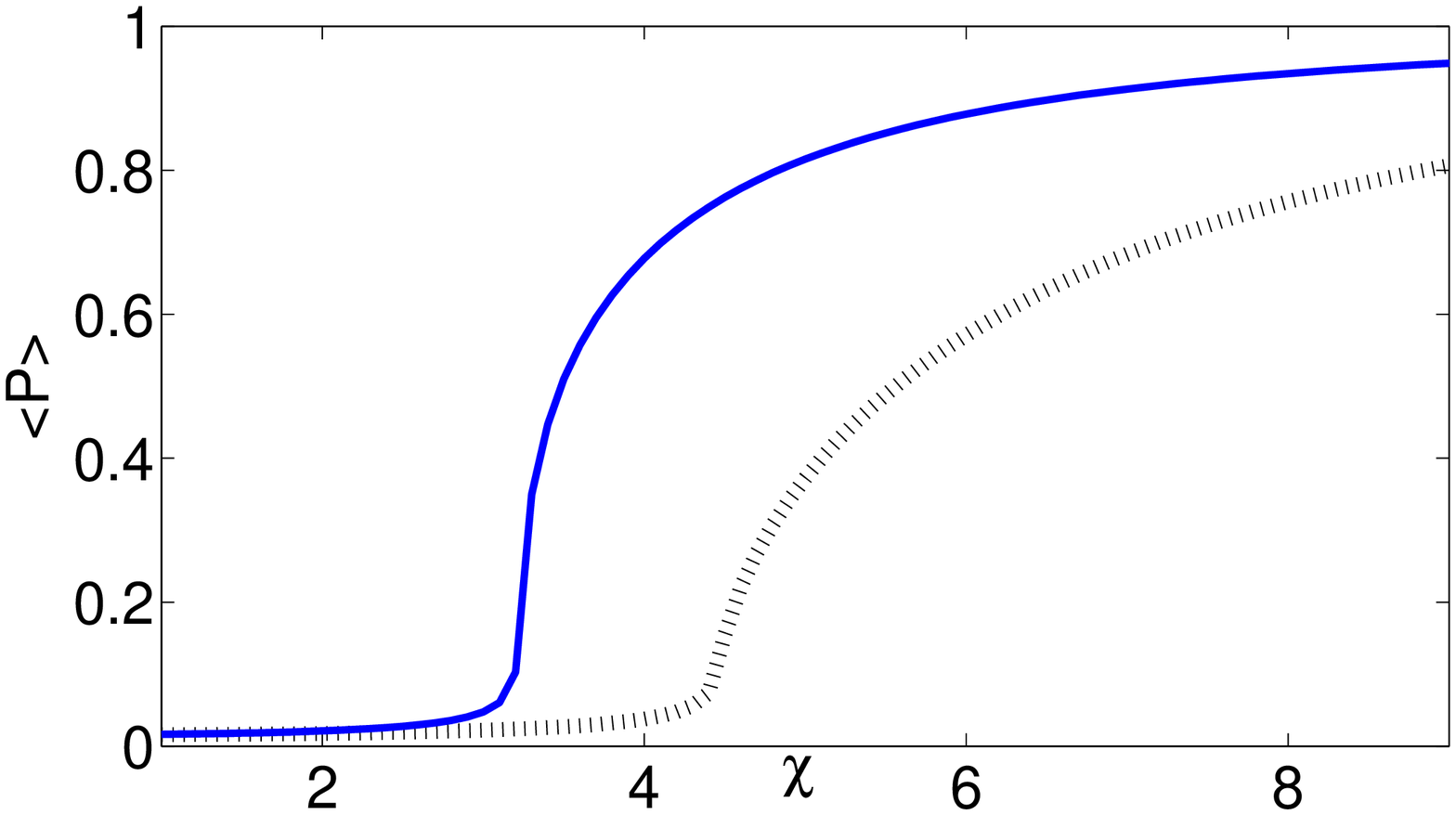}
    \end{tabular}
    \caption{(Color Online) Dependence of $\langle P \rangle$ on $\chi$.  
The top panel shows the comparison of the cubic
case (blue solid line) with the quintic case (red dashed line) and
the septimal case (green dash dotted line) for the power nonlinearity, while the bottom panel
contains the comparison of the cubic with the saturable (black
dotted line) case. In all cases, $V=1$.}
    \label{fig2a}
\end{center}
\end{figure}

In this section, we briefly revisit the relevant computations of
the quantities of interest such as $\langle P \rangle$ for 
signaling the relevant self-trapping dynamical transition in the $1$D-lattice, as well as extending the computations in the case of $2$D and $3$D-lattices.  For $V=1$ and 
for the cases of $\sigma=1$, $2$ and $3$, the relevant quantity
is shown (past an initial transient interval) for different values
of $\chi$ in the top panel of Fig.~\ref{fig2a}. The case of
$\sigma=1$ was also shown in~\cite{molina2} and 
in agreement with the computations of 
the latter, we find that $\chi_c \approx 3.2$ in this 
case. In the quintic case of $\sigma=2$, the corresponding value
becomes $\chi_c=5.48$, while for the septimal case of 
$\sigma=3$, the relevant critical point shifts to 
$\chi_c=7.05$. It is also interesting to point out that
the relevant curves become progressively steeper, as we 
increase $\sigma$. 
We observe that in all cases the numerical $\chi_c$ satisfies the lower bound predicted by Theorem \ref{Thstates} and naturally also of Corollary \ref{ThstatesC}, for the formation of stationary states, i.e., $\chi_c>\chi_{\sigma, N, V}$.  
Moreover, in accordance with expectations suggested by Corollary \ref{CV2}, the relevant
critical point is, in fact, increasing as a function of
$\sigma$. 

The bottom panel of Fig.~\ref{fig2a} shows a similar comparison but now
with the saturable nonlinearity. The latter  appears
to possess a higher value of $\chi_c \approx 4.4$, and this is quite close to the critical value $\chi_{s,N,V}=4$, for the creation of stationary states, as predicted by Corollary \ref{ThstatesC}.B. 
We observe that the actual critical value  $\chi_c$ in the saturable case is higher than the one in the cubic case, and the ordering of the critical points  $\chi_{\sigma, N, V}<\chi_{s,N,V}$ for stationary states suggested by  Corollary \ref{ThstatesC} is also valid for the dynamical problem, when the cubic nonlinearity is considered.
However, in consonance to the expectation
of the saturable being 
a more proximal case to the linear one, the increase in the relevant
dependence of $\chi$ is less steep and occurs over a wider
interval of nonlinearity strengths.


\begin{figure}
\begin{center}
    \begin{tabular}{cc}
    \includegraphics[scale=0.35]{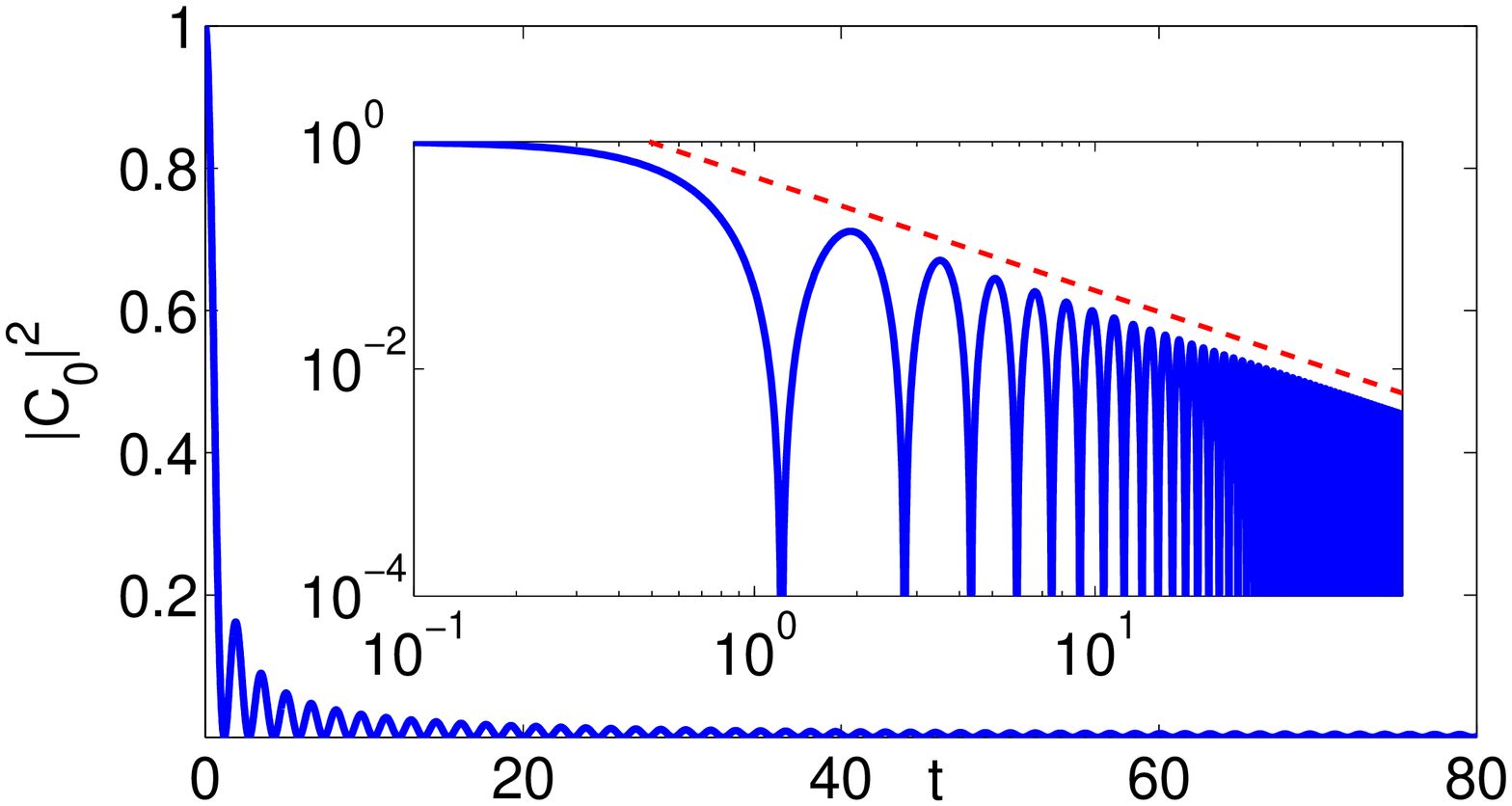}\\
    \includegraphics[scale=0.35]{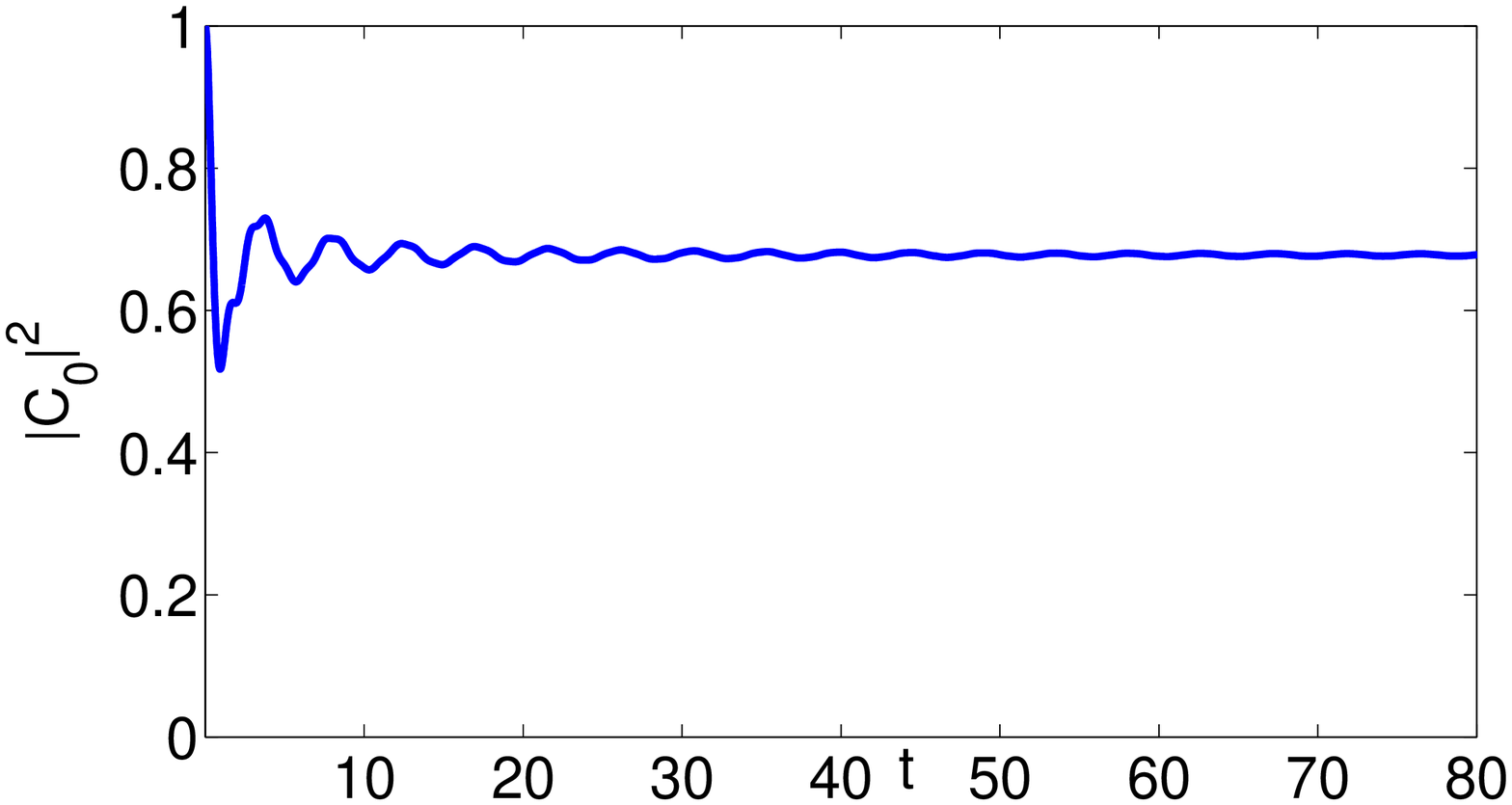}
    \end{tabular}
    \caption{(Color Online) Upper panel: typical example of the decay of the central site
square modulus $|C_0|^2$ as a function of time
(for $\chi=0.01$), in the case of the cubic power nonlinearity. The inset shows
the decay of $|C_0|^2$ over time in a log-log scale, with a $t^{-1}$ power
law given for comparison by the (red) dashed line. Bottom panel:
typical example of the convergence of the initial amplitude to a roughly
constant value for $\chi=4 > \chi_c$. In all cases, $V=1$.}
    \label{fig3a}
\end{center}
\end{figure}

\begin{figure}
\begin{center}
    \begin{tabular}{cc}
    \includegraphics[scale=0.34]{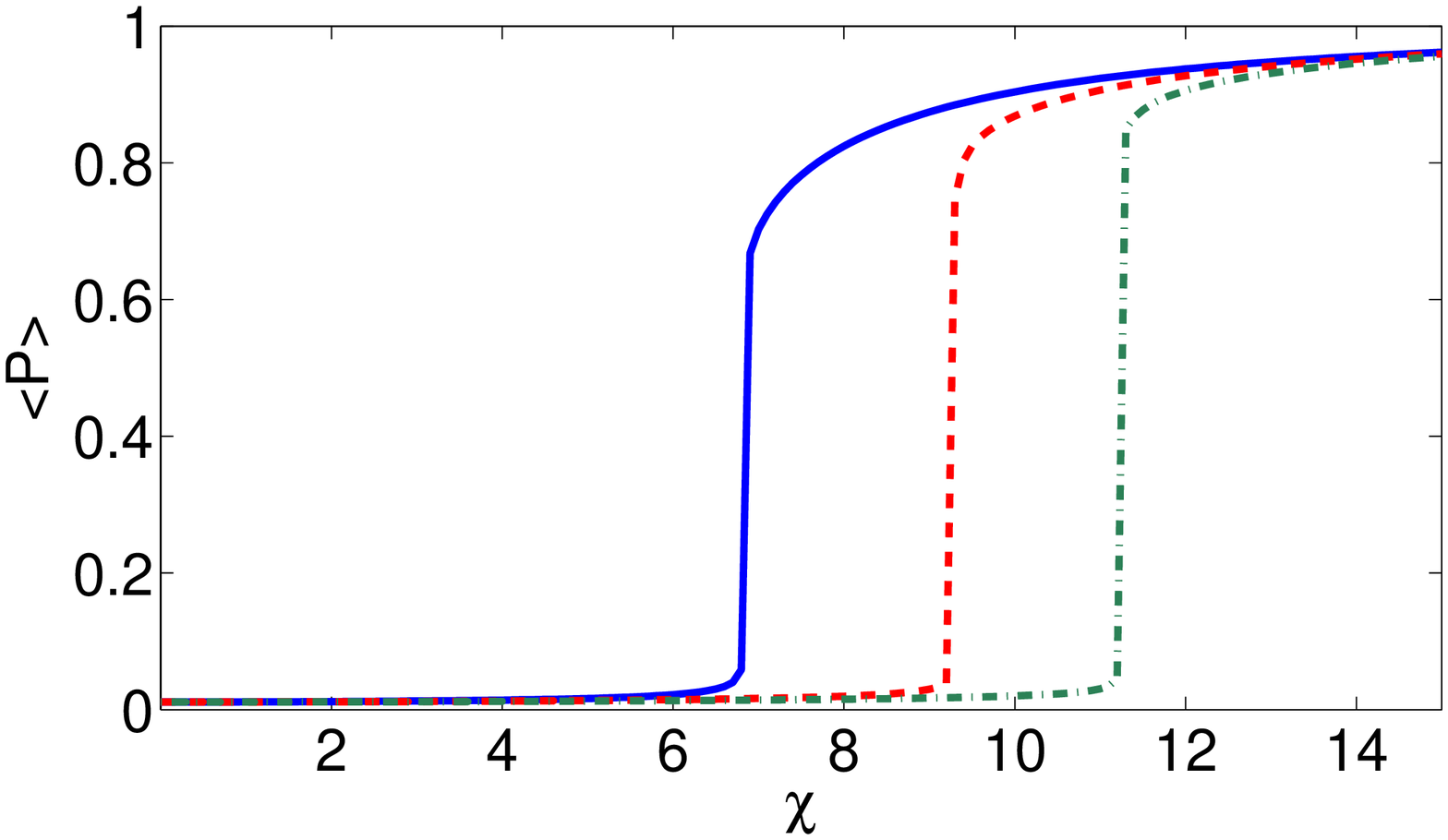}\\
    \includegraphics[scale=0.34]{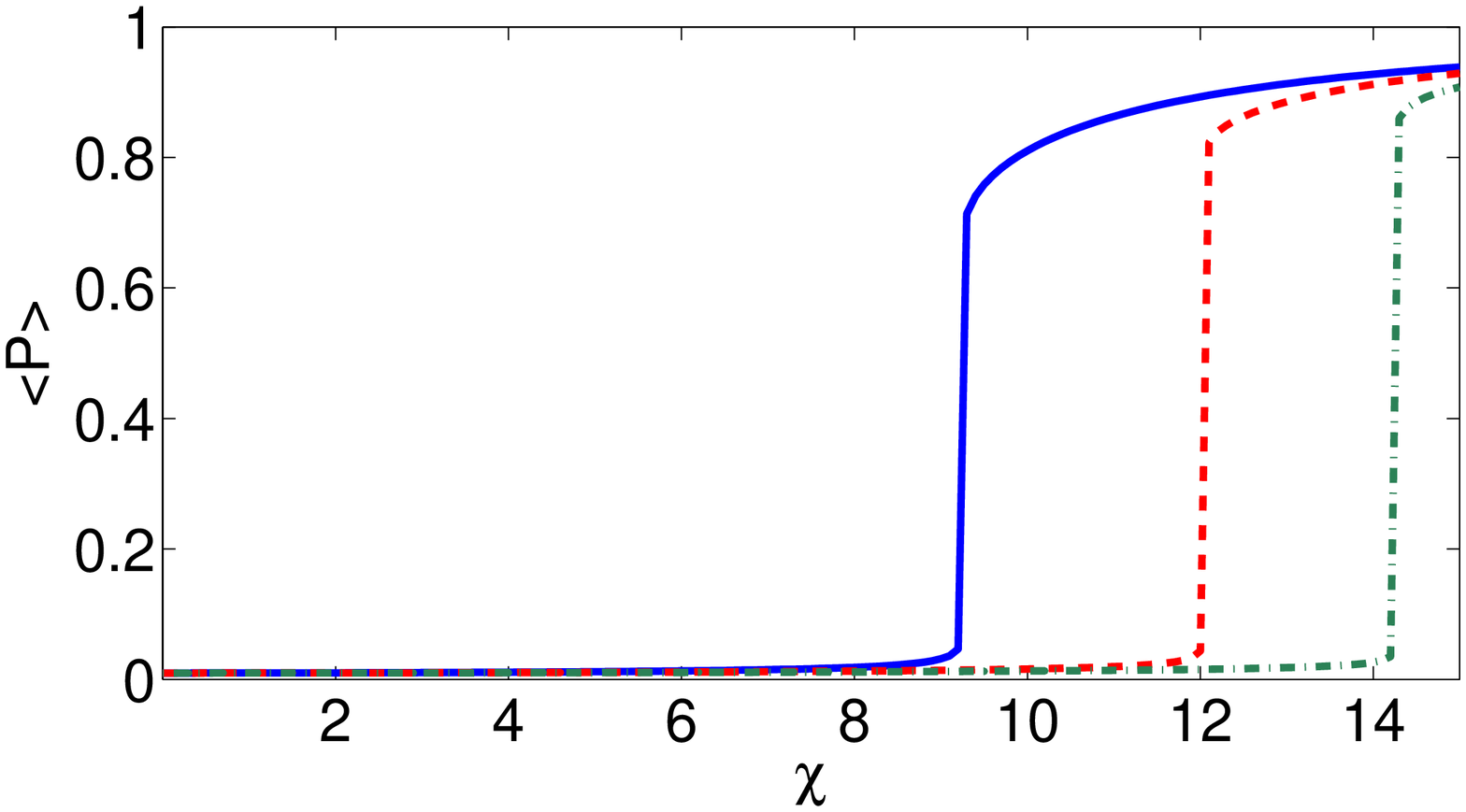}
    \end{tabular}
    \caption{(Color Online) Dependence of $\langle P \rangle$ on $\chi$ in  $2$D and $3$D-lattices, in the case of the power nonlinearity.
    The top panel shows the comparison of the cubic
    case (blue solid line) with the quintic case (red dashed line) and
    the septimal case (green dash dotted line) for  lattice dimension $N=2$, while the bottom panel
    shows the comparison for lattice dimension $N=3$.  In all cases, $V=1$.}
    \label{fig2D3D}
\end{center}
\end{figure}

It is  also interesting to note that the decay dynamics follows
a very typical pattern similar to the one shown in the upper 
panel of Fig.~\ref{fig3a}. The corresponding log-log plot inset
indicates that the central site amplitude decreases according to
a $t^{-1}$ power law (in terms of its envelope).
Essentially, in accordance with our
arguments in Sec. II (recall (\ref{eq8lin}) and the relevant discussion above), once the linear regime sets in the nonlinearity is irrelevant and plays a negligible role in the
ensuing dynamics, which is governed by the linear decay. 
 On the other hand,
as illustrated in the right panel, past the actual critical
point $\chi_c$, only a transient decay is observed, past
which the amplitude settles to a constant value and to the
corresponding defect mode. 
The quantity $\langle P \rangle$ in the case of $2$D and $3$D-lattices is shown in the upper and lower panel of Fig.~\ref{fig2D3D}, respectively for the case of the power nonlinearity.  The  numerical critical values $\chi_c$ when $\sigma = 1,2,3$, in the $1$D, $2$D and $3$D cases, are given in table \ref{Table1}. Remarkably, we observe the considerable increase of the critical values with respect to the dimension, while as a function of the nonlinearity exponent $\sigma$, the real critical point $\chi_c$ preserves its monotonicity property as in the $1$D-case. Both properties have been rigorously predicted for the value $\chi_{\sigma,N, V}$ for the creation of stationary states by Corollaries \ref{ThstatesC}.A and \ref{CV2}.1-2. The monotonicity property was also expected for the dynamical problem, at least in the case of the $3$D-lattice, from our asymptotic linearization arguments in Sec. II-see eq. \ref{monp2}. The last column of table \ref{Table1}, justifies that the theoretical value $\chi_{\sigma, N,V}$ is also of reasonable quantitative value as an approximation to the lower bound of $\chi_c$ for the dynamical self-trapping transition.  

\begin{figure}
\begin{center}
    \begin{tabular}{cc}
    \includegraphics[scale=0.34]{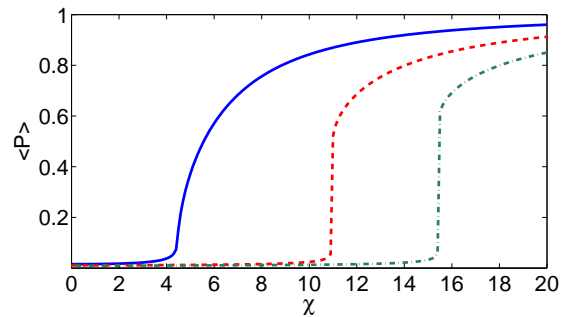}
    \end{tabular}
    \caption{(Color Online) Dependence of $\langle P \rangle$ on $\chi$ in  $1$D, for $2$D and $3$D-lattices, in the case of the saturable nonlinearity.
    The figure summarizes the comparison of the $1$D
    case (blue solid line) with the $2$D case (red dashed line) and
    the $3$D case (green dash dotted line). In all dimensional cases, $V=1$.}
    \label{figAllDsat}
\end{center}
\end{figure}

Fig. \ref {figAllDsat} summarizes the plots of $\langle P \rangle$ against $\chi$  for the saturable nonlinearity. We observe again the considerable increase of the critical values with respect to the dimension of the lattice, as predicted by 
Corollary \ref{CV2}.1.  Also, table \ref{Table2} highlights that the theoretical threshold $\chi_{s, N,V}$ for the creation of stationary states is again of reasonable quantitative value as a prediction of the lower bound of $\chi_c$ for 
the dynamical self-trapping transition.   

\begin{figure}
\begin{center}
    \begin{tabular}{cc}
    \includegraphics[scale=0.35]{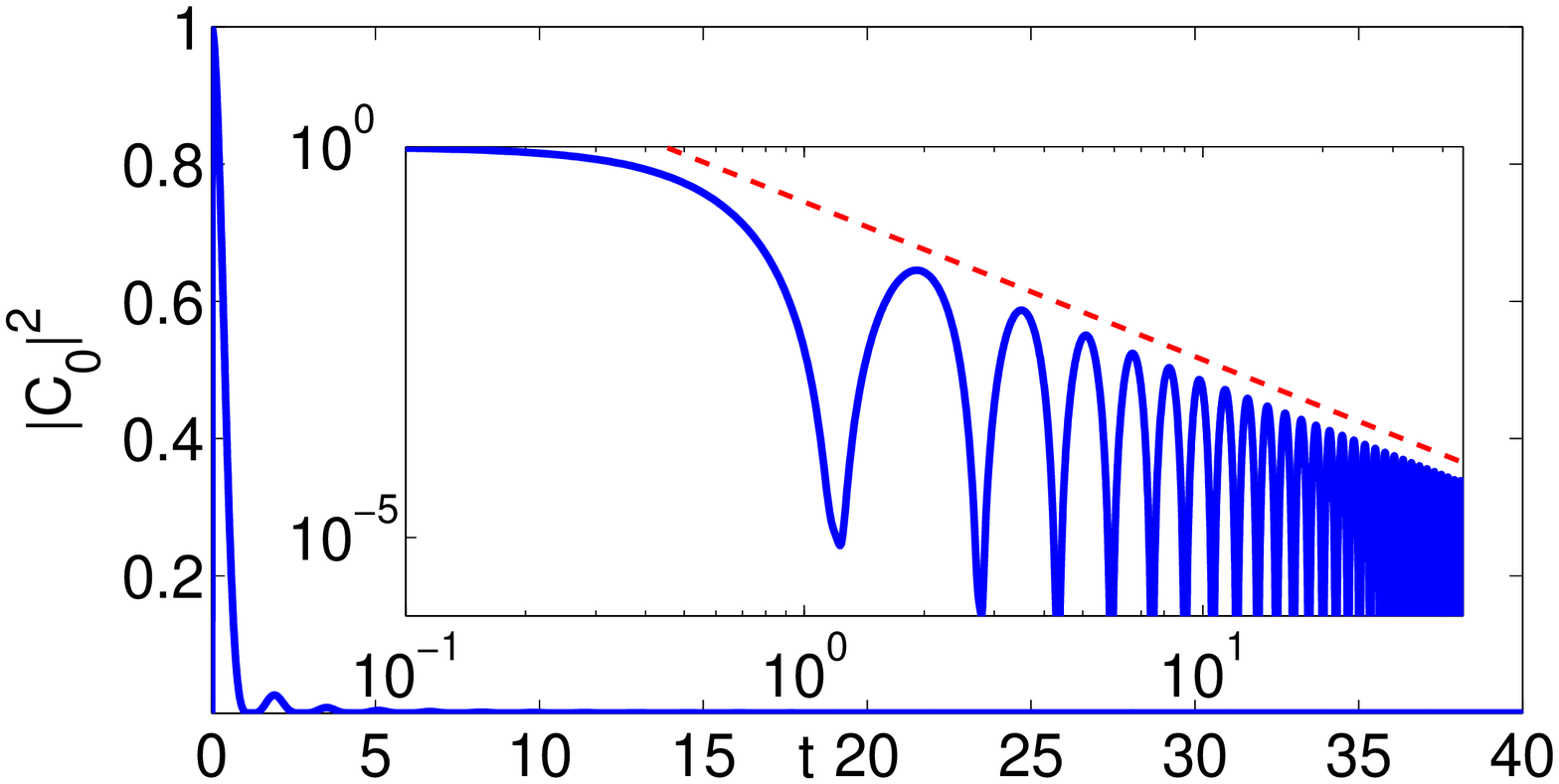}\\
    \includegraphics[scale=0.35]{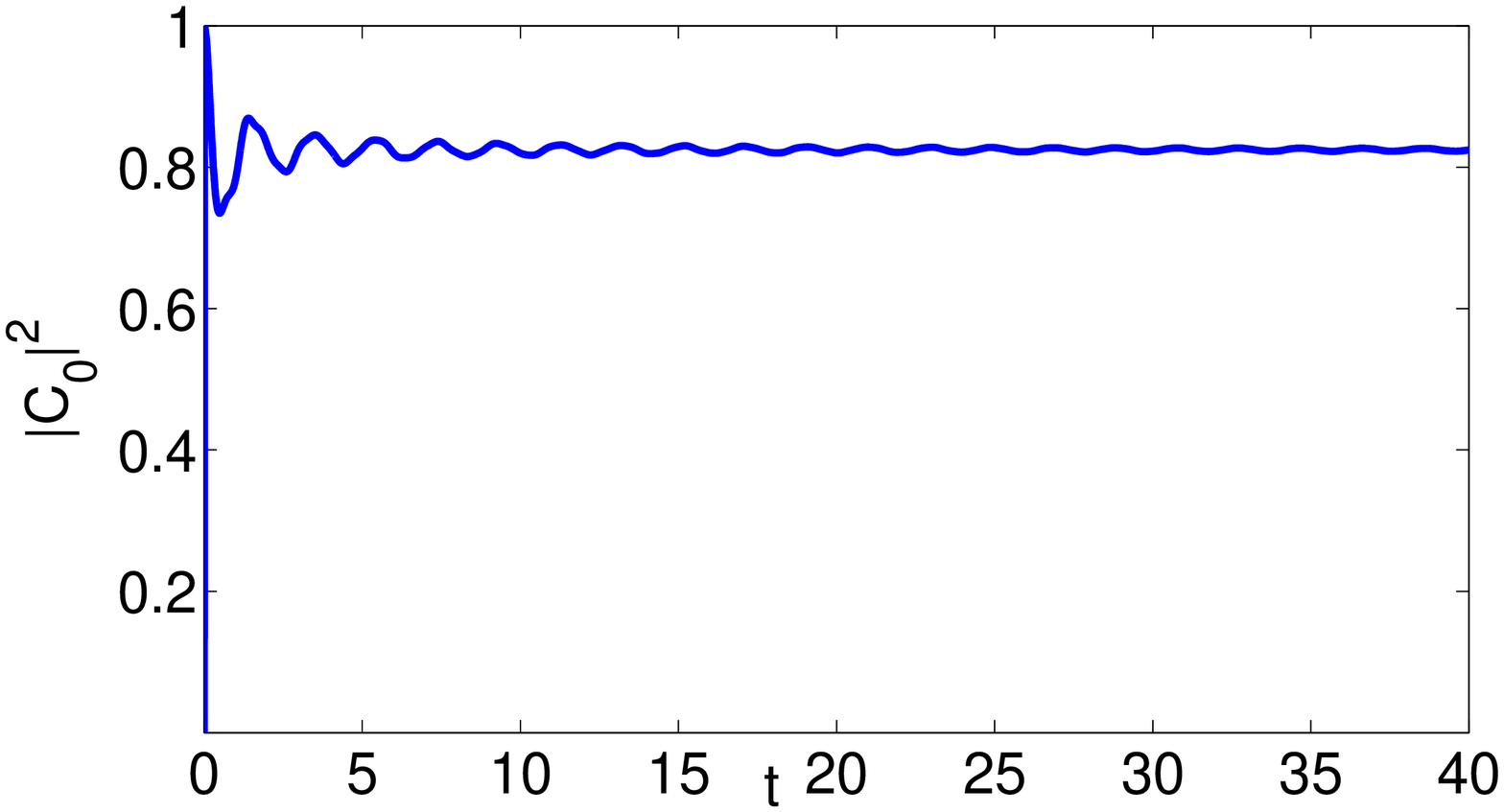}
    \end{tabular}
    \caption{(Color Online) Decay and self-trapping  in the $2$D-lattice, in the case of the cubic power nonlinearity. Top panel: typical example of the
    decay of the central site square modulus $|C_0 |^2$ as a function
    of time (for $\chi$ = 0.1). The inset shows
    the relevant decay in a log-log scale, with a $t^{-2}$ power
    law given for comparison by the (red) dashed line. Bottom panel: typical example of
    the convergence of the initial amplitude to a roughly constant
    value for $\chi = 8 > 6.8 = \chi_c$ . In both cases, $V = 1$, $\sigma=1$.  }
    \label{figC2D3Da}
\end{center}
\end{figure}

\begin{figure}
\begin{center}
    \begin{tabular}{cc}
    \includegraphics[scale=0.35]{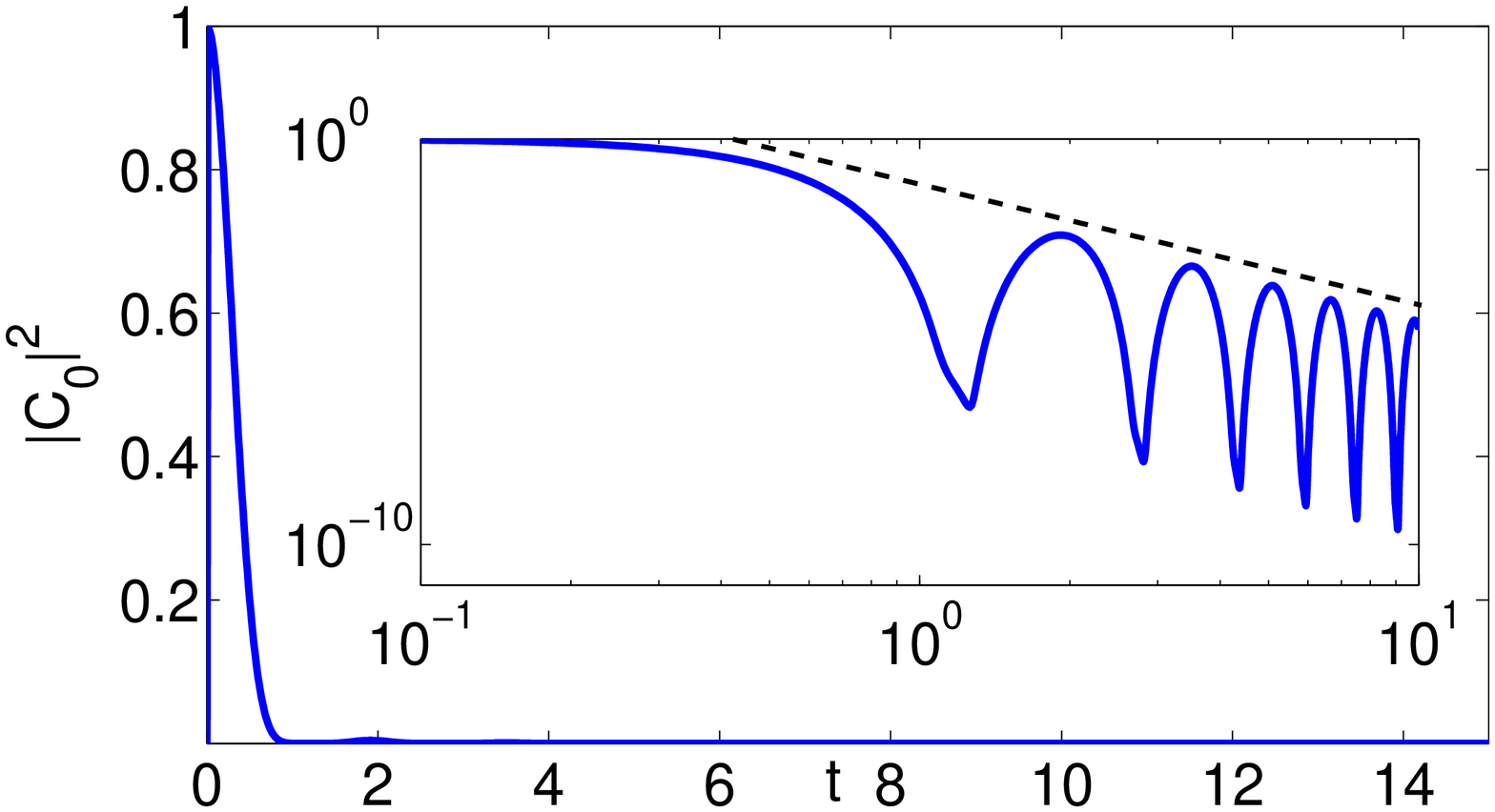}\\
    \includegraphics[scale=0.35]{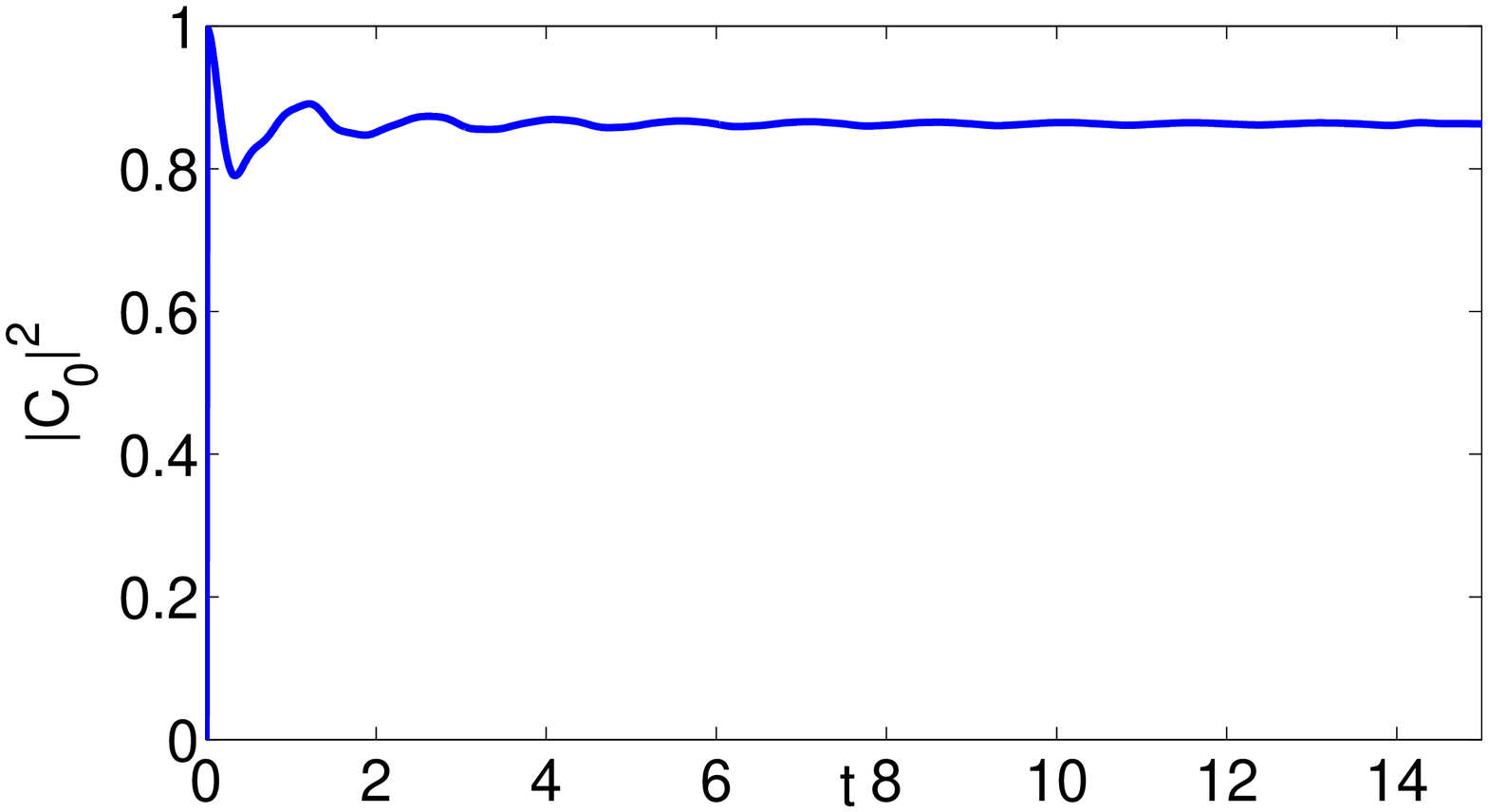}
    \end{tabular}
    \caption{(Color Online)  Decay and self-trapping in the $3$D-lattice, in the case of the cubic power nonlinearity. Top panel: typical example of the
    decay of the central site square modulus $|C_0 |^2$ as a function
    of time (for $\chi$ = 0.1). The inset shows
    the relevant decay in a log-log scale, with a $t^{-3}$ power
    law given for comparison by the (red) dashed line. 
    Bottom panel: typical example of
    the convergence of the initial amplitude to a roughly constant
    value for $\chi = 11 > 9.2 = \chi_c$ . In all cases, $V = 1$, $\sigma =1$. }
    \label{figC2D3Db}
\end{center}
\end{figure}

Figs. \ref{figC2D3Da}  and \ref{figC2D3Db} show typical examples of the decay and self-trapping dynamics in the $2$D and $3$D-lattices, respectively. The examples concern the power nonlinearity.  In the examples of the weak nonlinearity regime shown in the upper panels of  Figs. \ref{figC2D3Da}  and \ref{figC2D3Db}, we observe a drastic decrease of the  time interval for the transient, non-vanishing behavior, and an almost instantaneous decay. This effect is even stronger in the $3$D-case. On the other hand, in the examples of the strong nonlinearity regime shown in the bottom panels, we observe a decrease of the time interval of the transient behavior before the convergence of the amplitude to its constant value, and to the corresponding defect mode.  This behavior  combined with the increase of the critical points with respect to dimension, suggests the following: for higher dimensional lattices, stronger nonlinearity effects should be considered for self-trapping and localization of energy, due to a potential energy dispersion and delocalization associated with the higher dimensional set-ups. 

\begin{figure}
\begin{center}
    \begin{tabular}{cc}
   \includegraphics[scale=0.35]{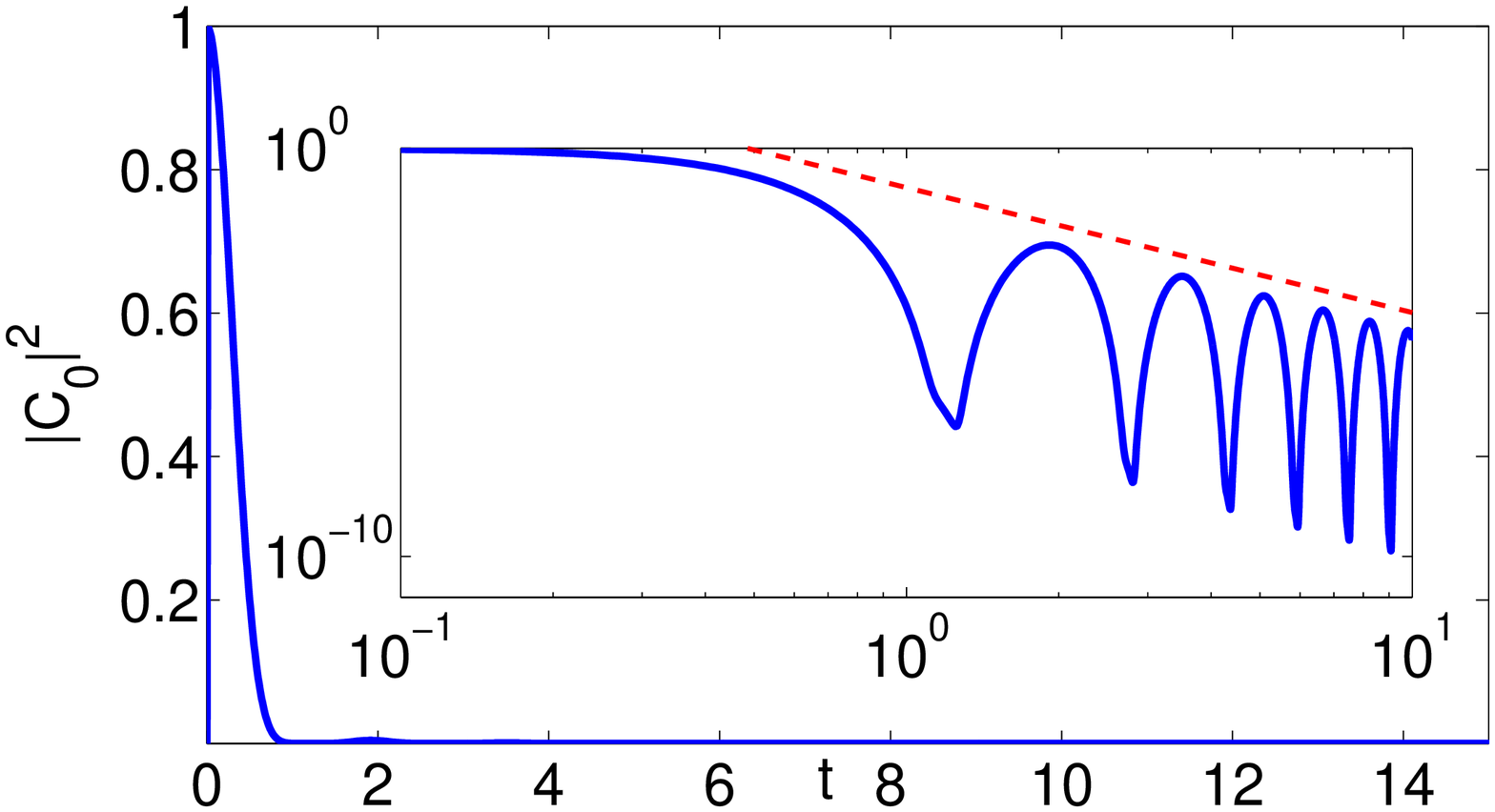}\\
    \includegraphics[scale=0.35]{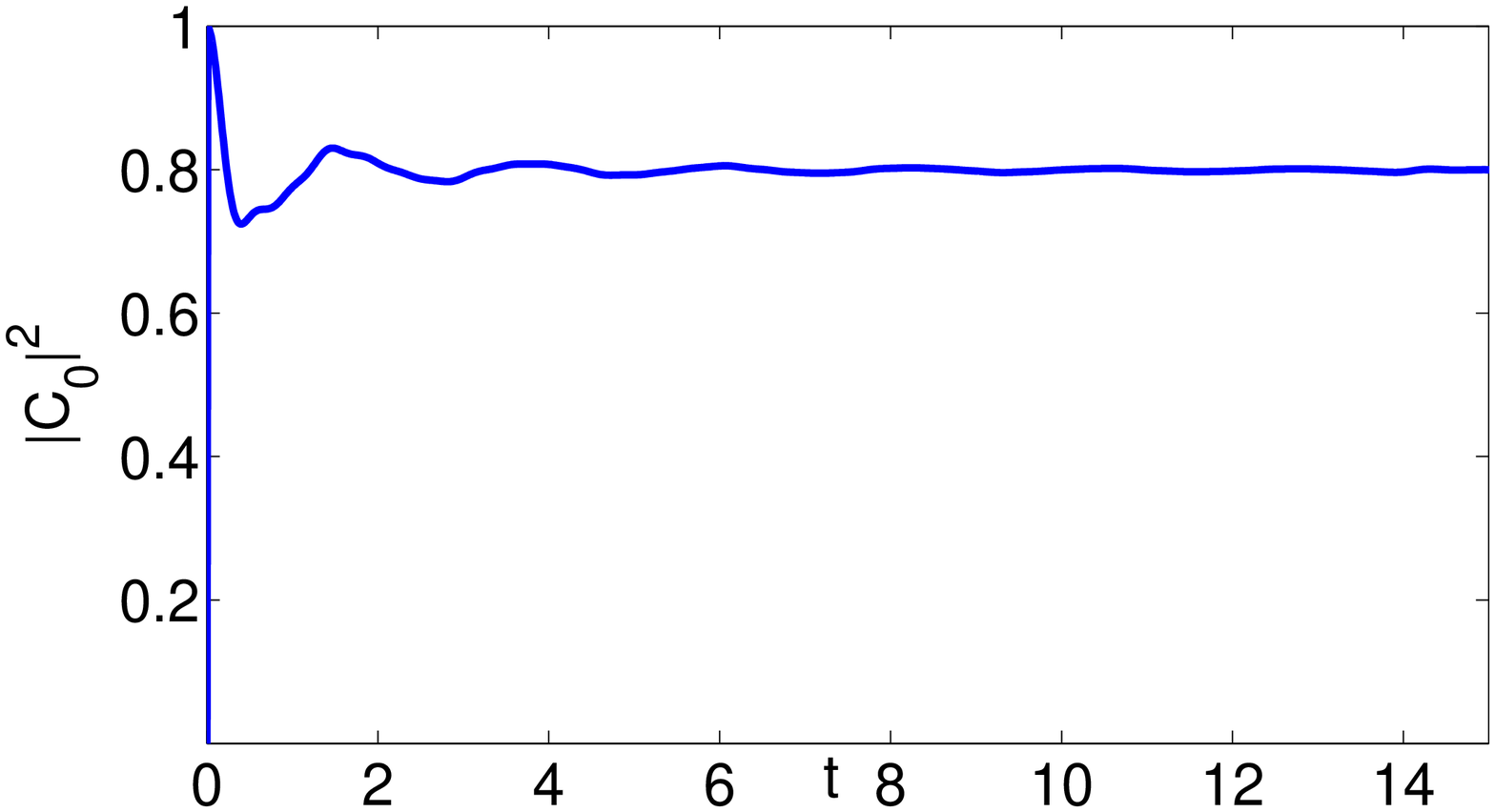}
    \end{tabular}
    \caption{(Color Online)  Decay and self-trapping in the $3$D-lattice, in the case of the saturable nonlinearity. Upper panel: typical example of the
    decay of the central site square modulus $|C_0 |^2$ as a function
    of time (for $\chi$ = 0.1). The inset shows
    the relevant decay in a log-log scale, with a $t^{-3}$ power
    law given for comparison by the (red) dashed line. Bottom panel: typical example of
    the convergence of the initial amplitude to a roughly constant
    value for $\chi = 18 > 15.4 = \chi_c$ . In all cases, $V = 1$, $\sigma =1$. }
    \label{fig3Dsat}
\end{center}
\end{figure}

Similar features were observed in the case of saturable nonlinearity. A typical example is shown in Fig.~\ref{fig3Dsat}. In particular, in the bottom panel concerning the strong nonlinearity regime, we observe a rapid convergence of $|C_0|^2$ to the stationary mode, similarly to what is shown in 
the $3$D example for the cubic power nonlinearity (shown in the bottom panel of Fig.~\ref{figC2D3Db}). 

We finally remark on the rate of decay of the central site amplitude in the $2$D and $3$D-cases.  The corresponding log-log plot insets in Figures \ref{figC2D3Da}, \ref{figC2D3Db} and \ref{fig3Dsat}, 
indicate that the central site amplitude decreases according to
a $t^{-N}$ power law (in terms of its envelope), for $N=2,3$, in the case of the power nonlinearity. Thus for all $1\leq N\leq 3$, the numerical results suggest that $|C_0(t)|^2\sim t^{-N}$. 
This could be reasonable by the linear approximation, where $|C_0(t)|^2 \sim |J_0(2Vt)|^{2N}$. Having in mind that $|J_0(x)|\sim\frac{1}{\sqrt{2\pi x}}$ for $x$ large, the linear approximation clearly suggests that $|C_0(t)|^2 \sim ct^{-N}$. Such a power law could associate the diagnostics (\ref{eq0}) and (\ref{eq0a}), in the sense that such a power law decay implies that $\langle P\rangle\sim 0$.

\begin{table}[b!]
\begin{tabular}{ |l| c| c| c| c|}
\hline
$N$ &  $\sigma=1$ & $\sigma=2$ & $\sigma=3$ & $\chi_{\sigma,N, V}$\\\hline
$1$  &  3.2 &   5.48&   7.05 & 2\\
\hline
$2$  &  6.8 &   9.2&   11.2 & 4\\
\hline 
$3$ &   9.2 &   12.0&   14.2 & 8\\\hline
\end{tabular}
\caption{Power law nonlinearity: the numerically found critical values $\chi_c$ for dynamical self-trapping in $1$D, $2$D and $3$D lattices. The last column shows the relevant critical value for the creation of stationary states $\chi_{\sigma,N, V}$, predicted by Corollary \ref{ThstatesC}.A. In all cases $V=1$.}
\label{Table1}
\end{table}
\begin{table}[b!]
\begin{tabular}{ |l| c| c|}
\hline
$N$ &  $\chi_c$  & $\chi_{s,N, V}$\\\hline
$1$  &  4.4  & 4\\
\hline
$2$  &  10.9 & 8\\
\hline 
$3$ &   15.4 & 12\\\hline
\end{tabular}
\caption{Saturable nonlinearity: the numerically found critical values $\chi_c$ for dynamical self-trapping in $1$D, $2$D and $3$D lattices. The last column shows the relevant critical value for the creation of stationary states $\chi_{s,N, V}$, predicted by Corollary \ref{ThstatesC}.A. In all cases $V=1$.}
\label{Table2}
\end{table}

\section{Conclusions and Future Challenges}

In the present paper, we have revisited the widely relevant theme 
of a single nonlinear defect embedded in an otherwise linear lattice.
We have addressed this problem from an up to now missing rigorous dynamical
perspective enabling the characterization of a weak nonlinearity
regime, via a suitable contraction mapping argument. This enabled
us (for different nonlinearity strengths and for different
nonlinearity forms --power law and saturable--) to come up with a 
proof of the fact that $\lim_{t \rightarrow \infty}|C_0(t)|^2=0$ for  a sufficiently
weak nonlinearity, at least in the $3$D-case. The technical restriction to dimension $N=3$ is imposed by the divergent nature of the integrals of the kernel of the associated integral equation, in lower dimensions $N=1,2$.

On the other hand, motivated by numerical observations that in the self-trapping regime the dynamics suggest convergence to a stationary mode, we have considered the existence of minimal values of the nonlinearity strength for the creation of such modes.  Our purely variational approach, enabled the construction of such modes as minimizers of the Hamiltonian energy, suggesting their dynamical stability as ground states. Their existence was proved under the assumption that the nonlinearity strength satisfies explicit lower bounds, relevant to those which may be obtained by local bifurcation theory, and in particular, bifurcation from the principal eigenvalue. Interestingly enough, the lower bounds for the creation of stationary states share the main qualitative features of the actual 
critical values for self-trapping observed in the dynamical problem, namely: the increase of the critical value as dimension increases, and the increase of the critical value as the nonlinearity exponent of the power nonlinearity also increases. The former suggests that stronger nonlinear effects are needed  to compensate energy dispersion in higher-dimensional lattices.  Moreover, comparing the critical values for the cubic power nonlinearity against the critical values for the saturable one, it was found that higher amplitudes are needed in the saturable case for self-trapping, than in the cubic case.  This ordering of critical values can be explained by the sub-linear (e.g. closer to the linear limit) nature of the saturable nonlinearity.  Notably, this effect is also suggested by the variational bounds of critical points for the creation of stationary modes.    
Naturally, there are numerous open questions that are still in need
of rigorous analysis. An important one concerns the extension of the rigorous dynamical arguments in the lower dimensional cases (dimensions of the lattice $N=1,2$). Another important question concerns an improved estimation of the true threshold separating the weakly nonlinear (dispersing) 
from the self-trapping regime.  However, clearly the techniques to
address the latter issue should be of a fundamentally different
kind than the ones used herein, and could be possibly based on analytical bifurcation theory for nonlinear integral equations.  On a related note, generalization
of  either the former (contraction mapping, possibly combined with invariance arguments for the integral map in suitable function spaces) 
or the latter (analytical bifurcation theory) considerations to multi-dimensional
or multi-defect settings would also be of particular interest.
However, there remain intriguing questions even at the computational
level and even in the one-dimensional setting. For instance, 
a separate theme of interest for future study would be 
to examine the asymptotic 
profile to which the self-trapping dynamics results and the ``selection 
principle''  of this state, among the family of possible stationary states
with different norms/energies. Similar investigations have been
numerically initiated also in the case of the fully nonlinear DNLS
lattice in~\cite{drossinos}.
These themes are presently under consideration and will be reported
in future publications.
\vspace{2mm}

{\it Acknowledgments}. PGK gratefully acknowledges support from
the NSF under grants DMS-1312856 and CMMI-1000337, as well as from 
the AFOSR under grant FA950-12-1-0332, the
Binational Science Foundation under grant 2010239, from the
Alexander von Humboldt Foundation and the FP7, Marie 
Curie Actions, People, International Research Staff 
Exchange Scheme (IRSES-606096). 
He also acknowledges the hospitality of
the IMA of the University of Minnesota during the final stages of
this work. MIM acknowledges support from Fondo Nacional de 
Ciencia y Tecnolog\'{\i}a (Grant 1120123), Programa Iniciativa Cient\'{\i}fica Milenio (Grant P10-030-F) and Programa 
de Financiamiento Basal (Grant FB0824).
\section{Appendix}
\setcounter{equation}{0}
\subsection{Eigenvalues of the Dirichlet Discrete Laplacian}
\label{AppPa}
In this part of the complementary section we provide for completeness, some information on the eigenvalue problem for the  discrete Laplacian. 
For
$n:=(n_1,n_2,\ldots,n_N)\in\mathbb{Z}^N$, we consider the problem
\begin{eqnarray}
\label{DLap}
-V\Delta_d\phi_n=\lambda\phi_n,
\end{eqnarray}
considered in the $N$-dimensional cube of $\mathbb{Z}^N$ with edges of length $L$,
\begin{eqnarray*}
&&\overline{\mathcal{Q}}=\{(x_{n_1},\ldots,x_{n_N})\,:\,0\leq n_1,\ldots,n_N\leq K+1\},\\
&&x_{n_i}=-L+n_{i}h,\;\;h=\frac{L}{K+1},\;\;i=1,\ldots,N.
\end{eqnarray*}
The discrete eigenfunctions on $\overline{\mathcal{Q}}$ are denoted by $$\phi_n=\phi(x_{n_1},x_{n_2},\ldots,x_{n_{N}}).$$
The interior of the cube $\overline{\mathcal{Q}}$ is given by
\begin{eqnarray*}
\mathcal{Q}=\{(x_{n_1},\ldots,x_{n_{N}})\,:\,1\leq n_1,\ldots,n_N\leq K\},
\end{eqnarray*}
and (\ref{DLap}) is supplemented with Dirichlet boundary conditions
\begin{eqnarray}
\label{DLapB}
\phi_n=0,\;\mbox{on}\;\partial\mathcal{Q}:=\overline{\mathcal{Q}}\setminus\mathcal{Q}.
\end{eqnarray}
Thus, the finite dimensional problem (\ref{DLap})--(\ref{DLapB}) is
formulated in the finite dimensional subspace of the sequence space $\ell^2$, 
\begin{eqnarray}
\label{subs}
\ell^2(\mathbb{Z}^N_K)=\left\{\phi\in\ell^2\;:\;\phi_n=0\;\;\mbox{for}
  \;\;||n||>K\right\}.
\end{eqnarray}
where $||n||=\max_{1\leq i\leq N}|n_i|$. 

When $\phi_n\in\mathbb{C}$, the space $\ell^2(\mathbb{Z}^N_K;\mathbb{C})$ becomes a real Hilbert space, $\ell^2(\mathbb{Z}^N_K;\mathbb{C})\equiv\ell^2(\mathbb{Z}^N_K;\mathbb{R})\times \ell^2(\mathbb{Z}^N_K;\mathbb{R})$, if endowed with the real inner product 
\begin{eqnarray*}
(\phi,\psi)_2=\mathrm{Re}\sum_{||n||\leq K}\phi_n\bar{\psi}_n.
\end{eqnarray*}
In the above real Hilbert space which will be still denoted for simplicity, by $\ell^2(\mathbb{Z}^N_K)$, the operator $-V\Delta_d:\ell^2(\mathbb{Z}^N_K)\rightarrow \ell^2(\mathbb{Z}^N_K)$ is self-adjoint.  Moreover, the eigenvalues of the discrete eigenvalue problem (\ref{DLap})-(\ref{DLapB}) in the finite dimensional space $\ell^2(\mathbb{Z}^N_K)$, coincide with those of its real counterpart  (e.g. for $\phi_n\in\mathbb{R}$), and are the following:
\begin{eqnarray*}
\lambda_{(n_{1},n_{2},\ldots,n_{N})}&=&4V[\sin^2\left(\frac{n_1\pi}{4(K+1)}
  \right)+\sin^2\left(\frac{n_2\pi}{4(K+1)}\right)\\
  &&+\ldots+
  \sin^2\left(\frac{n_N\pi}{4(K+1)}\right)],\\
n_i&=&1,\ldots,K\;\;i=1,\ldots,N,
\end{eqnarray*}
while its  principal eigenvalue is 
$$
\lambda_1\equiv\lambda_{(1,1,\ldots,1)}=
4VN\sin^2\left(\frac{\pi}{4(K+1)}\right).
$$
According to the variational characterization of the eigenvalues of the discrete Laplacian in the finite dimensional subspaces $\ell^2(\mathbb{Z}^N_K)$, 
$\lambda_1>0$, can be characterized as
\begin{eqnarray}
\label{eigchar}
\lambda_1=\inf_{
\begin{array}{c}
\phi \in \ell^2(\mathbb{Z}^N_K) \\
\phi \neq 0
\end{array}}\frac{(-V\Delta_d\phi,\phi)_{2}}{\sum_{|||n|||\leq K}|\phi_n|^2}.
\end{eqnarray}
Then, (\ref{eigchar}) implies the inequality
\begin{eqnarray}
\label{crucequiv}
\lambda_1\sum_{|||n|||\leq K}|\phi_n|^2&\leq&
(-V\Delta_d\phi,\phi)_{2}\nonumber\\
&\leq& 4V N \sum_{|||n|||\leq K}|\phi_n|^2.
\end{eqnarray} 
\subsection{Algebra with special functions}
\label{AppPb}
The integral in (\ref{N23}), is of the form
\begin{eqnarray}
\label{INK}
\int_{0}^{\infty} J_{0}(x)^2 \sin(A x) dx.
\end{eqnarray}
To calculate this integral, we recall first the relation of the Legendre function of the 1st kind $P_{\mu}(x)$ with the hypergeometric function $\mathcal{F}(a,b;c;x)$, for $x=\cos\phi$ \cite[Sec. 8.82-8.83, eq. (6)]{GR},
\begin{eqnarray}
\label{N25}
P_{\mu}(x)&=&P_{\mu}(\cos\phi)\nonumber\\
&=&\mathcal{F}\left(-\mu,\mu+1;1;\sin^2\frac{\phi}{2}\right)\nonumber\\
&=&\mathcal{F}\left(-\mu,\mu+1;1;\frac{1-x}{2}\right).
\end{eqnarray}
Setting $\mu=-1/2$ and $x=1-2A^2$ in (\ref{N25}) we get that
\begin{eqnarray}
\label{N26}
P_{-1/2}(1-2A^2)=\mathcal{F}\left(\frac{1}{2},\frac{1}{2};1;A^2\right).
\end{eqnarray}
We recall further, that the right hand side of (\ref{N26}) is associated the complete elliptic integral $K[A^2]$ of the 1st kind, with modulus $A$, \cite[Sec. 8.113, eq. (1)]{GR},
\begin{eqnarray}
\label{N27}
\mathcal{F}\left(\frac{1}{2},\frac{1}{2};1;A^2\right)&=&
\frac{2}{\pi}K[A^2]\\
&=&\frac{2}{\pi}\int^{\frac{\pi}{2}}_0\frac{d\phi}
{\sqrt{1-A^2\sin^2\phi}},\;\;0\leq A\leq 1\nonumber\\
&=&\frac{2}{\pi A}\int^{\frac{\pi}{2}}_0\frac{d\phi}
{\sqrt{1-\frac{1}{A^2}\sin^2\phi}}, A\geq 1.\nonumber
\end{eqnarray}  
Besides, from \cite[Sec. 6.672, eq. (5)]{GR},
\begin{eqnarray}
\label{N28a}
&&\int_{0}^{\infty} J_{\nu}(x)^2 \sin(A x) dx\nonumber\\
&=&\frac{1}{2}P_{\nu-1/2}(1-2A^2),\;\; 0<A<1,\;\;\mathrm{Re}\nu>-1,\nonumber\\
&=&\frac{1}{\pi}Q_{\nu-1/2}(2A^2-1),\;\;A>1,\;\;\mathrm{Re}\nu>-1,
\end{eqnarray}
where $Q_{\mu}$ is the Legendre function of the 2nd kind.  Setting $\nu=0$, in (\ref{N28a}), we get that
\begin{eqnarray}
\label{N28}
&&\int_{0}^{\infty} J_{0}(x)^2 \sin(A x) dx\nonumber\\
&=&\frac{1}{2}P_{-1/2}(1-2A^2),\;\; 0<A<1,\nonumber\\
&=&\frac{1}{\pi}Q_{-1/2}(2A^2-1),\;\;A>1.
\end{eqnarray}
Note that both functions $P_{\mu}$ and $Q_{\mu}$ are related for $x=\cos\phi$, with the formula \cite[Sec. 6.672, eq. (5)]{GR}, 
\begin{eqnarray}
\label{N29}
Q_{\mu}(-x)=-Q_{\mu}(x)\cos\mu\pi-\frac{\pi}{2}P_{\mu}(x)\sin\mu\pi.
\end{eqnarray}
Setting $\mu=-1/2$ and $x=1-2A^2$,  for $A>1$ in (\ref{N29}), we recover
\begin{eqnarray}
\label{N30}
Q_{-1/2}(2A^2-1)=\frac{\pi}{2}P_{-1/2}(1-2A^2),\;\;A>1.\;\;
\end{eqnarray}
Combining the branch of (\ref{N28}) for the case $A>1$, with (\ref{N30}), we find that
\begin{eqnarray}
\label{N31}
&&\int_{0}^{\infty} J_{0}(x)^2 \sin(A x) dx\nonumber\\
&=&\frac{1}{2}P_{-1/2}(1-2A^2),\;\; A>1.
\end{eqnarray}
Hence, from (\ref{N26}), (\ref{N27}) and (\ref{N31}) we conclude in 
\begin{eqnarray}
\label{N32}
&&\int_{0}^{\infty} J_{0}(x)^2 \sin(A x) dx\nonumber\\
&=&\frac{1}{\pi}\int^{\frac{\pi}{2}}_0\frac{d\phi}
{\sqrt{1-A^2\sin^2\phi}},\;\;0<A<1\\
&=&\frac{1}{\pi A}\int^{\frac{\pi}{2}}_0\frac{d\phi}
{\sqrt{1-\frac{1}{A^2}\sin^2\phi}}, A> 1.\nonumber
\end{eqnarray}


\end{document}